\documentclass[twoside]{article}
\usepackage[latin1]{inputenc}
\usepackage{mathptmx}
\usepackage{setspace}
\usepackage{t1enc}
\usepackage{times}
\usepackage{a4}
\usepackage{tabularx}
\usepackage{natbib}
\usepackage{graphicx}

\begin{document}
\setcounter{figure}{0}
\setcounter{section}{0}
\setcounter{equation}{0}

\begin{center}
{\Large\bf
High-(Energy)-Lights\\[0.2cm]
The Very High Energy Gamma-Ray Sky}\\[0.7cm]

Dieter Horns \\[0.17cm]
Institute for Experimental Physics, University of Hamburg\\
Luruper Chaussee 149, D-22761 Hamburg\\
dieter.horns@desy.de
\end{center}

\vspace{0.5cm}

\begin{abstract}
\noindent{\it
The high-lights of ground-based very-high-energy (VHE, $E>100$~GeV)
gamma-ray astronomy are reviewed. The summary covers both Galactic and extra-galactic sources. 
Implications for our understanding of the non-thermal Universe are discussed. Identified VHE 
sources include various types of supernova remnants (shell-type, mixed morphology,
composite) including pulsar wind nebulae, and X-ray binary systems. A diverse population of
VHE-emitting Galactic sources include 
regions of active star formation (young stellar associations), and massive 
molecular clouds.
Different types of active galactic nuclei have been found to emit VHE gamma-rays: besides predominantly Blazar-type objects, 
a radio-galaxy and a flat-spectrum radio-quasar have been discovered. Finally, many (presumably Galactic) 
sources have no convincing counterpart and remain
at this point unidentified. A total of at least 70 sources are currently known.
The next generation of ground based gamma-ray instruments aims to cover the entire accessible
energy range from as low as $\approx 10$~GeV up to $10^5$ GeV and to improve the sensitivity by 
an order of magnitude in comparison with current instruments.
}
\end{abstract}

\section{Introduction}
\begin{figure}
\begin{center}
\includegraphics[angle=90,width=1.00\linewidth]{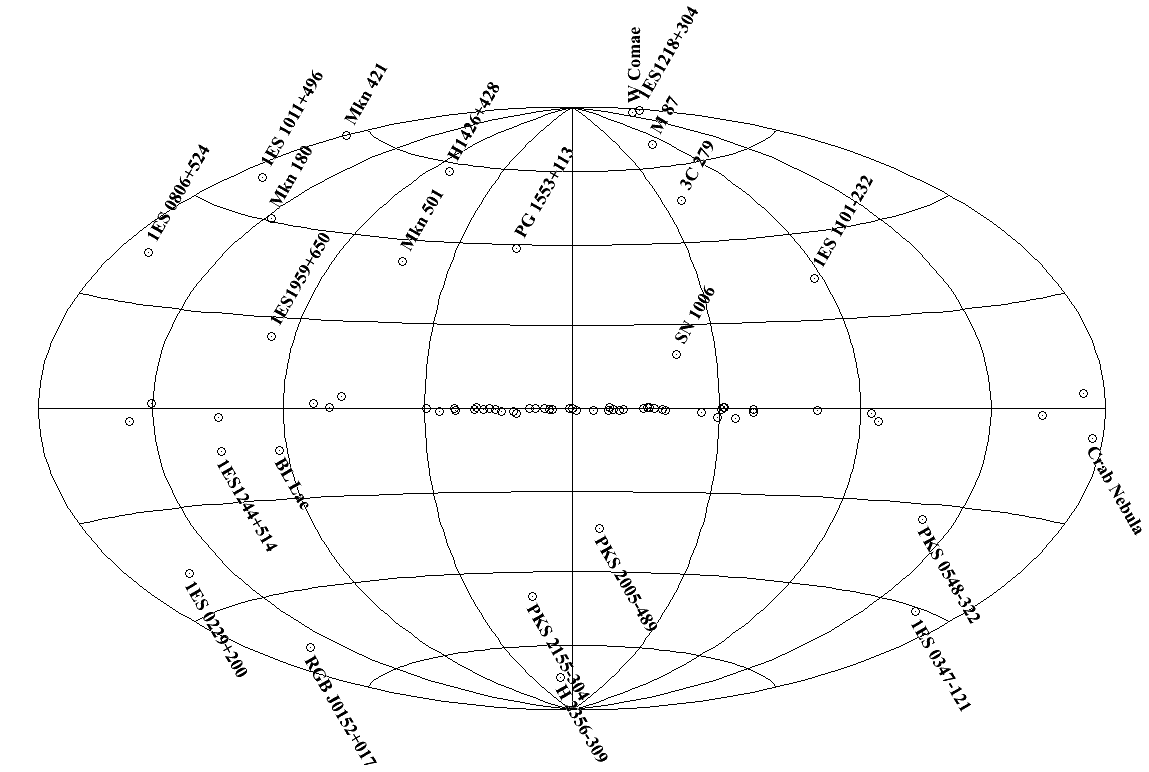}
\end{center}
\caption{\label{fig:sky}
 All VHE sources as known in September 2008 are shown as circles in a Hammer-Aitoff
projection of a Galactic coordinate system. 
The Galactic  center is in the center of the picture.
}
\end{figure}

The highest energy photons known are produced in astrophysical processes 
involving even more energetic particles presumably accelerated through stochastic acceleration mechanisms as
suggested initially by \citet{1949PhRv...75.1169F}. 
Therefore, observations of VHE photons provide a direct view of the astrophysical accelerators of charged
particles and allow to identify the individual sources of cosmic rays: VHE photons open our view
to the ``accelerator sky''. The
origin of the Galactic population of charged cosmic rays has remained since its discovery by V. Hess in 1912 until today a long-standing question of astro- and particle physics. A widely favored model on the origin
of cosmic rays assumes that diffusive shock acceleration takes place in the expanding blast waves of supernova remnants converting $10-20~\%$ of the kinetic energy into cosmic rays. Under this assumption,
Galactic supernova remnants provide sufficient power ($\approx 10^{41}$~ergs~s$^{-1}$) in order to balance the escape losses of Galactic cosmic rays as well as produce a power-law type distribution of particle energy that closely resembles the cosmic ray spectrum measured locally \citep[see e.g.][]{1964ocr..book.....G,2006astro.ph..7109H}.\\
Observations of VHE-gamma-rays, mainly by imaging air Cherenkov telescopes in the last
decade, have surpassed the anticipated detection of a few supernova remnants  
and have established a rich and diverse collection of VHE sources.
Specifically,  the currently active generation of imaging air Cherenkov telescopes 
(H.E.S.S.\footnote{http://www.mpi-hd.mpg.de/hfm/HESS/HESS.html}, 
MAGIC\footnote{http://wwwmagic.mppmu.mpg.de/}, and VERITAS\footnote{http://veritas.sao.arizona.edu/}) have fulfilled and by far exceeded
the expectations that were based upon the pioneering previous generation
of experiments: the results obtained in the
last years have shown that VHE emission is common to a variety of different
source types - not only shell-type SNRs. \\
The current view of the source distribution in the VHE sky is shown in Figure~\ref{fig:sky} where
all known sources  are displayed (status of September 2008). 
Note, the sensitivity achieved  varies greatly across the sky. The
best sensitivity is reached in the inner Galactic disk where a dedicated survey with the air Cherenkov telescopes of the H.E.S.S. experiment
has been performed (see
section~\ref{subsect:imaging}).\\
The most remarkable feature of the gamma-ray sky is not evident 
in this picture: each source shown is an accelerator of particles up to and beyond TeV energies (``accelerator sky''). 
A similar conclusion can not be drawn when looking at source populations detected in other wavelength bands. 
This makes the VHE band a sensitive window to detect non-thermal particle accelerators. 
Future neutrino telescopes will almost certainly have the potential to detect some of the brighter VHE 
sources (and discover new sources that are optically thick for VHE gamma-rays). 
The detection of high energy neutrinos is experimentally a challenging 
task and requires tremendous efforts (see e.g. the
Ice-CUBE neutrino telescope in the Antarctic ice). However, 
the observation of a neutrino source will ultimately demonstrate the presence 
of accelerated nuclei - largely model-independent. Different to the clear observational signature
in the neutrino channel, VHE gamma-rays are sensitive to accelerated electrons (through inverse Compton
scattering) as well as to accelerated nuclei (through neutral meson production and decay).

The scope of this article is limited to observational highlights obtained with ground based 
instruments and does not aim at presenting a complete review of the field 
of VHE astrophysics.
The paper is structured in the following way: In  section~\ref{section:obstech}, 
the experimental techniques of ground-based instruments (both imaging and non-imaging)
are described before continuing with the census of today's VHE sky in section \ref{section:census}. 
Section~\ref{sec:secphy} provides a  
a short overview on VHE gamma-rays as probes of the interstellar medium including 
Lorenz-invariance violating effects related to 
structure of space-time at Planck-scales. Finally, in section~\ref{section:future} this review is concluded with 
some comments on the 
future of the field.

\section{Observational techniques}
\label{section:obstech}
Non-thermal gamma-ray sources produce typically power-law type energy spectra in which the flux drops
with increasing energy. As a consequence of this,   
space based detection techniques are  limited by the small detection rate of photons at high energies
($E<10-100~$GeV).  
The detection of VHE ($E>100~$GeV) photons requires collection areas
that are substantially larger than the ${\cal O}(\mathrm{m}^2)$ of satellite based detectors. One approach towards
large collection areas are ground based observation techniques which however are limited to energies sufficiently 
high ($E>5$ GeV)  to
produce a detectable air shower.  \\
Ground based gamma-ray observations are always based upon air shower
detection techniques. High energy particles (photons, electrons, and 
nuclei) initiate extensive air showers in the atmosphere which acts as 
a natural calorimeter: secondary particles are detected either  
when they reach the ground or  by the
Cherenkov light produced in the atmosphere. 
At ultra-high energies, it becomes feasible to detect
fluorescence light as well as radio signals which are not discussed here (see
e.g. the contribution by J. H\"orandel at this conference).\\
Generally, the air shower detection technique relies on one of the following approaches:
\begin{itemize}
\item Shower-front sampling of particles 
\item Shower-front sampling of Cherenkov light
\item Imaging of the air shower using Cherenkov light.
\end{itemize}
The benefits (marked '+') and draw-backs ('-') of the shower-front sampling of particles and the imaging air
Cherenkov technique are mainly the following\footnote{Shower-front sampling of Cherenkov light is discussed seperately below.}:
\begin{itemize} 
\item Imaging technique: 
 \begin{description}
 \item[+] sensitivity ($\approx 10^{-13}$~ergs/(cm$^2$~s)$^{-1}$ at TeV-energies, 50~hrs exposure)
 \item[+] angular resolution (few arc minutes per event)
 \item[+] spectroscopy ($\Delta E/E\approx 15$~\%)
 \item[+] low energy threshold 
 \item[-] field of view (a few degrees, $\approx $~msrad)
 \item[-] duty cycle ($\approx 10$~\%)
 \end{description}
\item Shower front sampling of particles:
 \begin{description}
  \item[+] high duty cycle ($\approx 95~\%$)
  \item[+] large field of view ($\approx 2$~srad)
  \item[-] high energy threshold
  \item[-] sensitivity ($\approx 10^{-12}$~ergs~cm$^{-2}$~s$^{-1}$, 1~year exposure)
  \item[-] angular resolution ($0.3-0.7^\circ$)
  \item[-] energy resolution ($\Delta E/E\approx 30-100$~\%)
 \end{description}
 \end{itemize}
 The obvious complementarity of the two techniques justifies the further development of both of them in the future:
non-imaging techniques cover simultaneously a large field of view and can
be operated 24 hours a day. On the other hand, air Cherenkov telescopes
have a narrow field of view (a few milli steradians), reaching a supreme
sensitivity, but can only be operated in clear, dark nights. \\
In addition to the indirect ground based techniques discussed until now, high energy gamma-rays can be
detected above the atmosphere directly by pair-conversion detectors.
 With these detectors, events initiated by charged particles can be effectively suppressed by e.g. a scintillation veto shield. 
Obviously, this is not possible for ground based indirect detection techniques where a separation of gamma-ray and cosmic-ray
events has to be done using the information of the air shower itself (gamma-hadron separation). 
 \\
With the commissioning of the Fermi satellite\footnote{formerly known as GLAST} \citep{2007APh....28..422A},
ground-based gamma-ray astronomy will benefit from the simultaneous operation
of essentially a sensitive all-sky monitor operating at lower energies. The
sensitivity of Fermi is well matched by the sensitivity of ground-based experiments so 
that the two instruments will be able to provide detection or constraints on the
spectral shape across 6-7 orders of magnitude in energy. 
\subsection{Shower-front sampling techniques}
\paragraph{Particle arrays.}
 The experimental approach for the measurement of cosmic rays as well as the search for gamma-ray sources had been 
largely dominated by air shower arrays until the beginning of the 1990's. These arrays of particle detectors 
measure the arrival time of the shower front and the lateral particle density distribution. The total number of particles
in the air shower is used to reconstruct the total energy of the primary particle. 
The relative timing of the individual 
particle detectors in the array is used to determine the direction of the air shower (typical
angular resolutions of 1$^\circ$ can be reached) while both the timing and particle density measured on the ground are in principle
useful to discern electromagnetic
from hadronic air-showers (gamma-hadron separation). However, the gamma-hadron separation is naturally limited
by the filling factor\footnote{the fraction of the total surface covered with active detector surface} 
of the detector array which typically amounts to  less than $1$\%. 
An important development in the field
 was the innovation introduced by the MILAGRO collaboration \citep[see e.g.][]{2004ApJ...608..680A}: they use
a pond of water which is instrumented with photo-multiplier tubes (PMTs).
The PMTs detect 
Cherenkov light from the secondary particles entering the pond's water. This way, the filling factor
is increased to almost 100\% and the entire particle distribution is sampled by the detector. Using the
\textit{clumpiness} of the particle distribution as an indicator for
a hadronic air shower\footnote{for electromagnetic air showers, the lateral distribution is
smoother than for a hadronic shower which contains muons and sub-showers}, it has been possible to reach sufficient sensitivities to detect 
VHE gamma-ray sources.\\ Future projects along this direction aim at installing 
such a detector at high altitude, where the energy threshold can be  decreased to reach values
below 1~TeV. Parallel to the MILAGRO group, the ARGO collaboration
has developed a new type of air shower detector using resistive plate chambers 
to increase the filling factor. Their installation is already located at high 
altitude and is starting operation \citep{2007NuPhS.166...96A}. For a review of ground based
non-imaging detectors, see \citet{2007JPhCS..60....1L}.
\paragraph{Shower-front sampling with  Cherenkov light.}
The experimental technique used by non-imaging air Cherenkov detectors marks the transition from non-imaging to imaging observations. These detectors sample the arrival time and density of air Cherenkov photons using either open PMTs like
e.g. the THEMISTOCLE  experiment \citep{1990NuPhS..14...79F}, AIROBICC array \citep{1995APh.....3..321K} or
large reflecting surfaces as e.g.   heliostats provided by solar power plants. The latter approach has been pursued  by a number
of groups including the C.A.C.T.U.S. \citep{marleau05}, CELESTE \citep{2002NIMPA.490...71P} as well
as STACEE \citep{2005astro.ph..6613S}. While arrays of open PMTs retain the large field
of view of classical air shower sampling arrays, the heliostat arrays have a very small field of view but a substantially
smaller energy threshold reaching below 100~GeV as achieved e.g. by the CELESTE experiment . 
The potential for discovering faint sources is limited by the gamma-hadron rejection power of these shower sampling
instruments.
\subsection{Imaging air Cherenkov telescopes}
\label{subsect:imaging}
 The field of ground-based gamma astronomy has been largely driven by the 
remarkable results obtained with the imaging air Cherenkov telescopes (IACTs). Extensive
air showers emit in the forward-direction a beam of atmospheric Cherenkov light with an opening angle 
of $\approx 1^\circ$. This beam illuminates almost homogeneously a circular
region on the ground with a diameter of 200-300 m (depending on the altitude and inclination of the shower axis). 
An optical telescope pointing parallel to the shower axis and located within the illuminated footprint of the shower can make an image of the
air shower against the background light of the night sky, provided the camera  is 
sufficiently fast to integrate the short
Cherenkov flash of only a few nano-seconds.  The image provides
information on the original particle's energy, direction, and on its nature (nucleus
or photon). The reconstruction of these parameters is improved considerably when
more than one telescope is used in  a stereoscopic set-up where the telescopes are separated by 50-150~m
in order to provide a baseline for triangulating the atmospheric air shower.  The stereoscopic 
technique has become the nominal standard for all current and future installations.\\ 
The performance of today's telescopes is essentially characterized by the sensitivity to detect
VHE sources with an energy flux down to $10^{-13}$~ergs~(cm$^2$~s)$^{-1}$ in 50~hrs
of observation time; this corresponds to a minimum detectable luminosity of 
$L_\mathrm{min}\approx10^{31}$~ergs~s$^{-1}$~$(d/1~\mathrm{kpc})^2$ at a distance of 1~kpc or more suitable for extra-galactic objects
$L_\mathrm{min}\approx10^{41}$~ergs~s$^{-1}~(d/100~\mathrm{Mpc})^2$ at a distance of 100~Mpc. \\
The angular resolution of each reconstructed primary $\gamma$-ray is typically better than $6$~arc~min. The relative energy
resolution is comparably good  and reaches 
values of $\Delta E/E\approx 10-20$~\%. This is sufficient to 
detect and characterize spectral features like curvature or even lines from
e.g. self-annihilation of Dark matter particles. For a list of currently operating IACTs see e.g. \citet{2007arXiv0712.3352H}. 
\section{A census of today's VHE gamma-ray sky}
\label{section:census}
 The currently known list of VHE sources encompasses more than 70 sources (see Fig.~\ref{fig:sky}). 
The number of sources is modest in comparison with catalogues assembled
at lower energies.  However,
it is remarkable, how many different types of objects are actually 
emitting VHE gamma-rays and are therefore accelerators of multi-TeV particles. 
\\
Before discussing the different source types that have been detected, it is enlightening to list promising source types which  
have not been detected in the VHE band so far (the
possible reason(s) for non-detection listed in parenthesis):
\begin{itemize}
\item \textbf{Star burst galaxies (too faint):} The high star forming rate should lead ultimately to an enhanced
rate of supernova explosions. The upper limits on VHE emission from e.g. NGC~253 \citep{2005A&A...442..177A} are still higher than the 
expected flux \citep{2005A&A...444..403D}  but deeper observations (ca. 50~hrs)
 will eventually reach the required 
sensitivity to either detect a signal or put meaningful constraints on the models\footnote{The CANGAROO
collaboration published a claim for a detection from the starburst galaxy NGC~253 which was however later
retracted \citep{2007A&A...462...67I}.}.
\item \textbf{Gamma-ray bursts (too short, too far):} These extremely powerful explosions are transient events requiring 
fast turn-around times and/or wide-field-of-view instruments in order to capture these objects during their outbursts. 
No
convincing GRB detection has been claimed so far.\footnote{Some evidence for GRB detections have been published in the past \citep[see e.g.][]{1998A&A...337...43P,2000ApJ...533L.119A}.} 
The ultra-fast slewing MAGIC telescope has succeeded in observing
(but not detecting) a GRB while the outburst was still ongoing \citep{2006ApJ...641L...9A}. 
Non-imaging all-sky
survey instruments like MILAGRO are probably more likely to detect these transient events including
also the distinct class of short GRBs \citep[for constraints see e.g.][]{2007ApJ...666..361A}. The visibility of distant GRB events is limited due to absorption of
the energetic photons emitted at cosmological distances in pair-production processes with the optical-to-infra-red background light. An energy threshold well below 100~GeV is crucial to be able to observe GRBs at red-shifts $z>0.1$.   
\item \textbf{Clusters of galaxies (too extended, too faint):} 
So far, no group or cluster of galaxies has been detected in the VHE band. 
Clusters of galaxies confine effectively cosmic rays and are expected to produce VHE-emission via inelastic scattering
of nuclei with the intra-cluster gas \citep{1996SSRv...75..279V,2004A&A...413...17P}.
In addition to cosmic rays injected by normal galaxies and accelerated in large-scale shocks, AGN activity could also lead
to an additional contribution to the intra-cluster cosmic ray density \citep{2007MNRAS.382..466H}. Current upper limits do
not constrain severely the non-thermal population of cosmic rays in the clusters \citep{2007arXiv0708.1384D}. 
\item \textbf{Pulsars (early cut-off in energy spectrum):} 
Fast rotating neutron stars are expected to give rise
to acceleration within the magnetosphere \citep[for a review, see e.g.][]{2006csxs.book..279K}. 
This leads to pulsed 
emission (mainly from curvature radiation) that has been detected at least 
from six isolated pulsars up to GeV energies with the EGRET spark chamber detector on board
the Compton Gamma-Ray Observatory \citep{1999ApJ...516..297T}. However,
the energy spectra of pulsars are also expected to show a sudden cut-off at a few GeV which
has made them so far invisible to  ground-based telescopes with an energy threshold  around
100~GeV. Most searches for pulsed emission above 100~GeV have so-far not been successful in finding VHE emission from 
pulsars - consistent with the expectations \citep{2007A&A...466..543A, 2007ApJ...669.1143A}.
The MAGIC collaboration has recently reported the first detection of pulsed 
emission with ground based instruments above 25 GeV at
the level of 6.4$~\sigma$ \citep{2008ATel.1491....1T}.
Additionally, pulsed emission has been suggested to 
be produced via bulk Compton-scattering processes in the un-shocked pulsar wind-zone \citep{2000MNRAS.313..504B}. Non-detection of pulsed emission
from the Crab pulsar has been used to constrain the formation region of the ultra-relativistic wind \citep{2004ApJ...614..897A}. 
\end{itemize}
The breakdown of known VHE sources includes 50 Galactic objects: 
5 shell-type SNRs, 2 mixed-morphology SNRs, 2 composite SNRs, 20 pulsar-wind nebulae,
2 stellar associations, 4 X-ray binary systems, and roughly 18 sources without clear association to known objects. The remaining 23 objects
are of extra-galactic origin: 21 Blazars, one Fanaroff-Riley Type I (M~87), and one flat-spectrum radio quasar (3C279). 
 \subsection{Shell-type supernova remnants}
\begin{figure}
\mbox{
\parbox{0.50\linewidth}{\includegraphics[width=\linewidth]{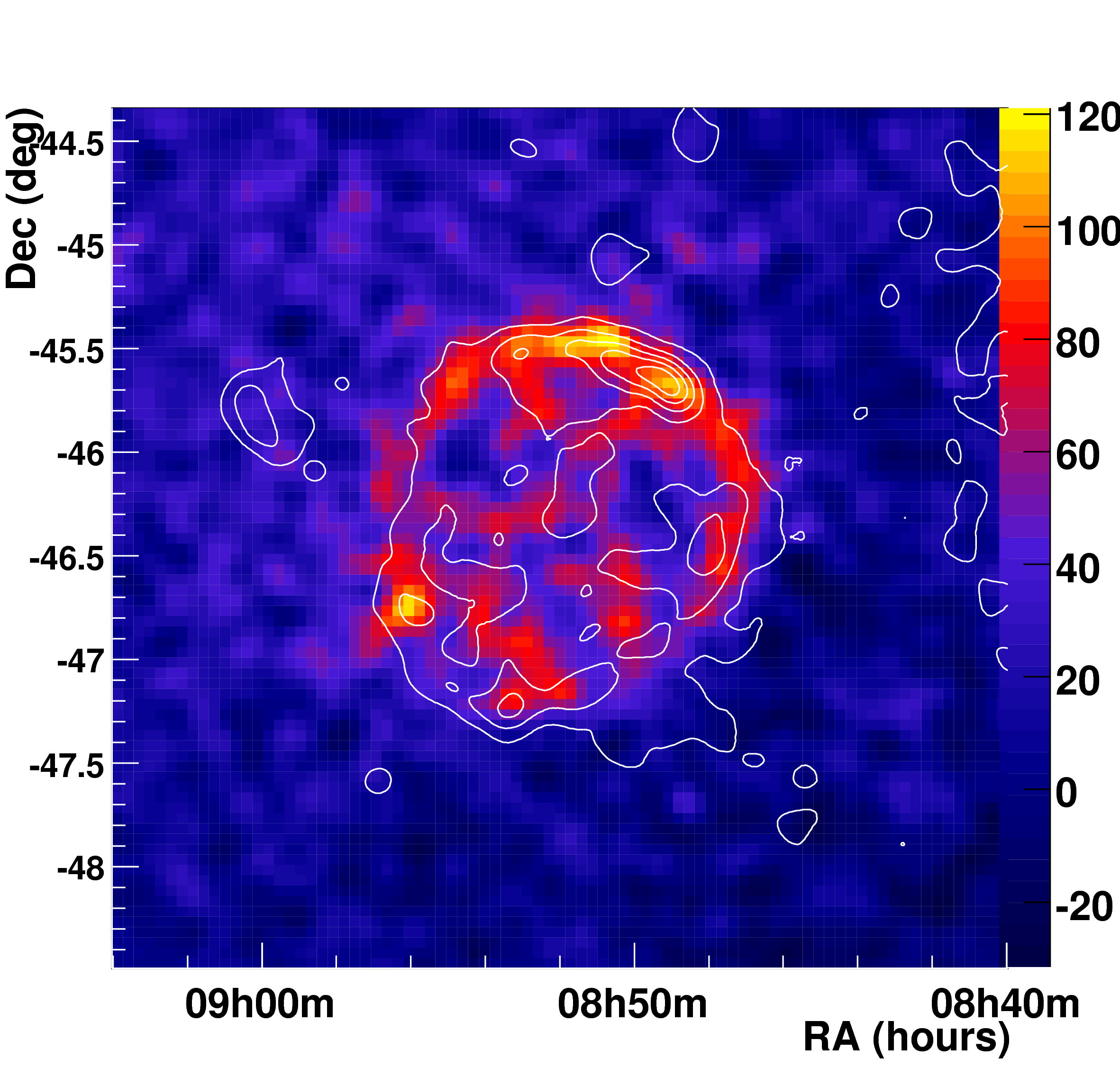}}
\parbox{0.50\linewidth}{\includegraphics[width=0.8\linewidth]{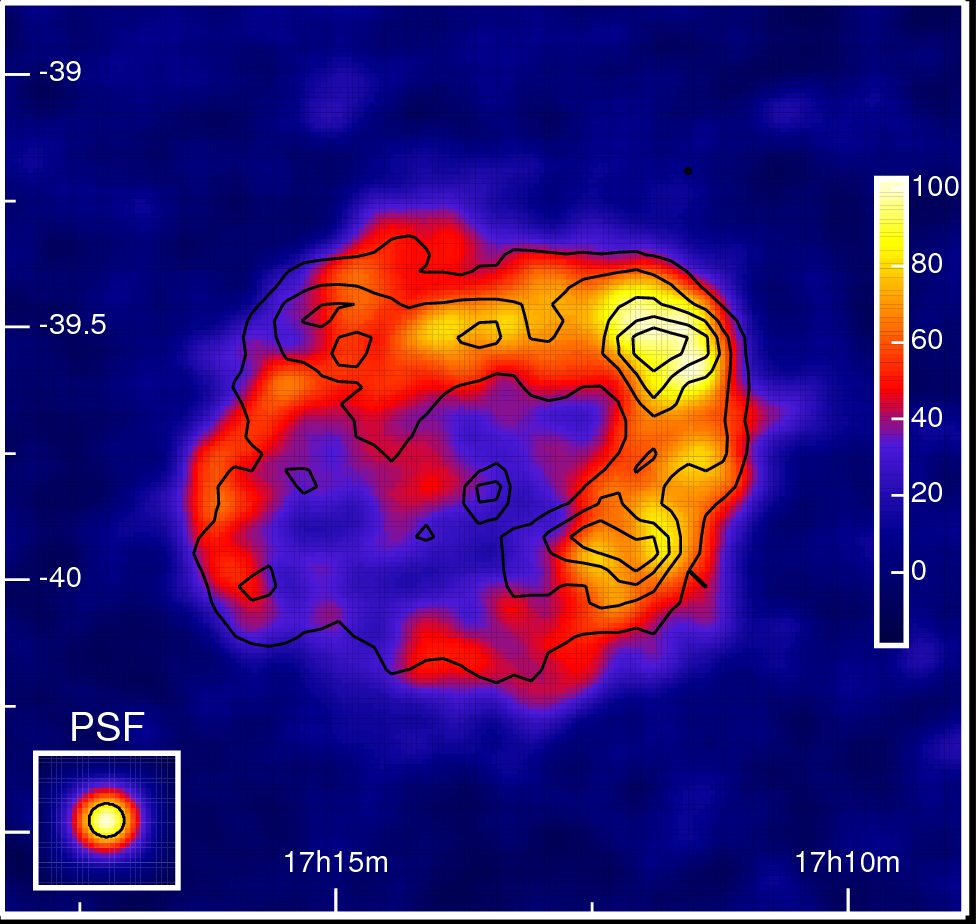}}}
\vspace{-0.9cm}
\mbox{
\hspace*{-0.4cm}
\parbox{0.50\linewidth}{\includegraphics[width=1.1\linewidth]{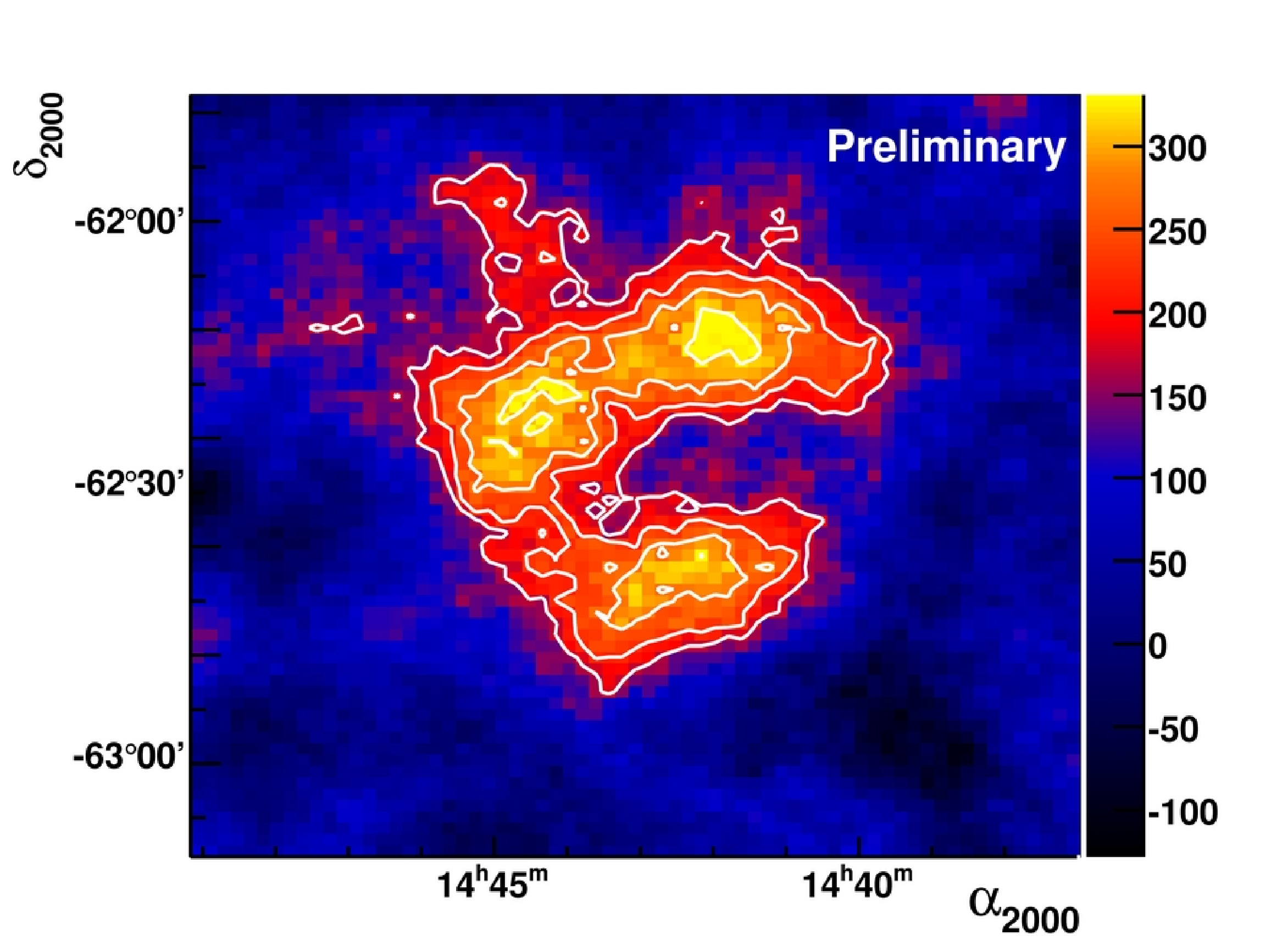}}
\parbox{0.50\linewidth}{\includegraphics[width=0.85\linewidth]{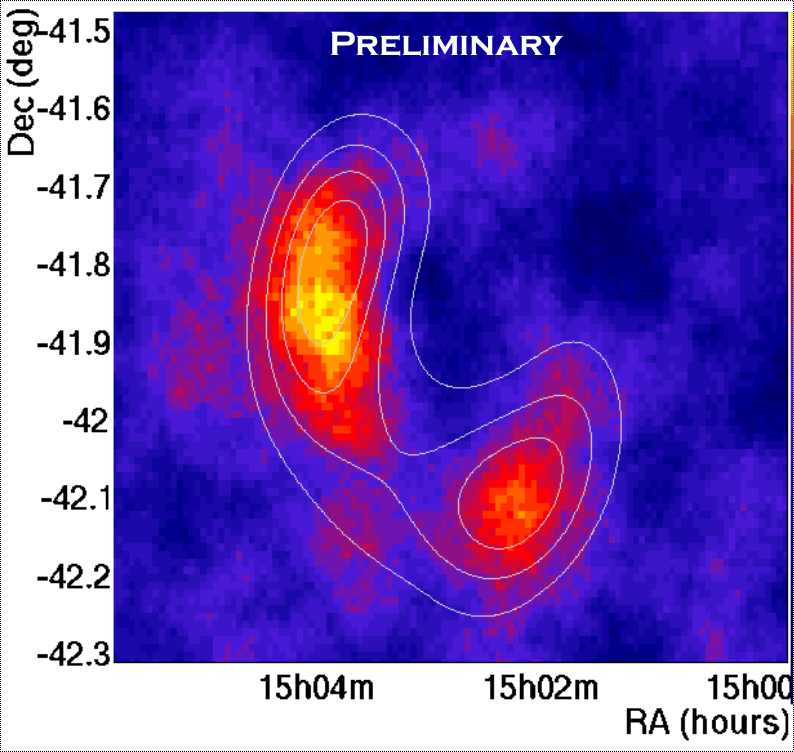}}}
\vspace{0.9cm}
\caption{\label{image:SNR} Upper panel, from left to right: RX~J0852.0-4622 ("Vela Jr") \citep{2007ApJ...661..236A}, 
RX~J1713.7-3946 \citep{2006A&A...449..223A}, bottom panel:  RCW~86 \citep{2007arXiv0709.4103H}, 
SN1006 \citep{2009AIPC..xxx..yyS}. The color-coded images
are the excess maps as obtained with the H.E.S.S. telescopes (see text for references). Overlaid in white contours (RX~J0852.0-4622) is the smoothed ROSAT image obtained above 1.3~keV. The black contours overlaid on the VHE-image of
RX~J1713.7-3946 follow the ASCA image (1-3~keV). The white contours overlaid on RCW~86 are the significance
contours (starting at 3 standard deviations, increasing by 1) The white contours on the preliminary excess map
from SN~1006 follows the Chandra X-ray map smoothed to match the H.E.S.S. point-spread function.}
\end{figure}
  Shell-type supernova remnants are commonly considered to be the best candidates to accelerate Galactic cosmic rays. During
  the supernova explosion (either a thermo-nuclear explosion or a core-collapse event), the stellar atmosphere
  is ejected with an initial velocity of a few $1000$~km~s$^{-1}$ carrying  $10^{51}$~erg in kinetic energy and driving
  a shock front in the interstellar medium. While slowing down 
  during the Sedov phase\footnote{once the SNR has swept up a comparable amount of matter from the interstellar medium 
  as its ejecta mass, the free expansion phase ends and the Sedov phase of SNR evolution starts}, the 
  expanding shock front heats up the ambient medium giving rise to thermal X-ray emission. The shock is expected to 
  dissipate a good fraction ($10-30$~\%) of its energy during the Sedov phase in the form of accelerated particles 
  (electrons and nuclei).
  In this case, the total power injected by roughly 1 supernova explosion every 30~years is sufficient to balance 
  the escape losses of Galactic cosmic rays.
  Radio- as well as X-ray synchrotron emission is detected and attributed to electrons accelerated by the 
  forward shock which in projection produces the characteristic shell-like morphology.   
  For a few objects, the non-thermal X-ray component dominates entirely the observed X-ray 
spectrum (e.g. SN~1006, RX~J1713.7-3946). \\
  However, radio and X-ray observations are sensitive mainly to electrons and only indirectly to the presence of 
energetic nuclei: 
X-ray observations with sub-arcsecond spatial resolution have revealed indirectly, that the forward shock 
in many SNRs is very likely modified by cosmic ray streaming instabilities leading to enhanced magnetic field
compression \citep{2000MNRAS.314...65L}. This in turn leads to  observable effects: 
  electrons produce  X-ray synchrotron emission in small regions with increased magnetic field upstream of the shock 
   that appear as narrow X-ray filaments ($d<1$~pc) that have been first detected using the superior angular 
   resolution of the Chandra X-ray telescope \citep{2003ApJ...589..827B}. 
  The inferred magnetic field strength in the X-ray filaments of e.g. SN1006 is of the order of 
  $100$~$\mu$G \citep{2003A&A...412L..11B} which is much larger than in the  case of a linear shock thus 
  indirectly revealing efficient acceleration of nuclei. Recent observations of 
  fast variability of X-ray emission in filamentary structures seems to provide additional evidence for strong 
   magnetic fields in RX~J1713.7-3946 \citep{2007Natur.449..576U}. 
\\
  VHE gamma-ray observations are sensitive to both, electrons as well as nuclei. Sufficiently energetic
electrons can radiate VHE gamma-rays through inverse Compton-up-scattering of soft seed photons 
 (leptonic origin of VHE-emission)
while nuclei produce neutral mesons decaying into gamma-rays in inelastic scattering events with the ambient gas 
 (hadronic origin of VHE-emission). \\
In the hadronic scenario, the energy loss-time for accelerated protons 
   ($\tau_{pp\rightarrow \pi^0}\approx 4.5\times 10^{15}~n^{-1}$~s) 
varies only slowly with
energy and  
depends mainly on the ambient 
medium density ($n_H=n$~cm$^{-3}$) which includes cold molecular and hot ionized gas as well as the ejecta of the
progenitor star. 
  The actual density of gas near the shock can be inferred e.g. by modeling thermal X-ray 
emission from heated and partially ionized gas. However, for SNR without thermal
  X-rays, the density can only be constrained rather loosely and uncertainties close to an order of magnitude can remain.
  The typical integrated luminosity in VHE gamma-rays (1-10~TeV) is from a population of energetic protons with total energy 
 $W_p$ $L_{\gamma}\approx W_p~\tau^{-1}=3.7\cdot 10^{33} \cdot \eta/0.1 \cdot (E_\mathrm{SN}/10^{51}~\mbox{ergs}) n$~ergs~s$^{-1}$ for an efficiency of shock acceleration $\eta=0.1$, ie. the fraction of the kinetic
energy of the blast wave converted into charged particles accelerated \footnote{The accelerated spectrum is assumed to follow 
a power-law with $dn/dE\propto E^{-2}$}.  A similar result on the SNR luminosity has 
 been obtained by \citet{1994A&A...287..959D}.\\
Currently four VHE-emitting shell-type SNRs
  are detected and upper limits have been derived for a few more objects (see Table~\ref{table:SNR} for a list). 
Except for Cassiopeia A, the spatial extension of the SNR has been resolved (see Fig.~\ref{image:SNR}). 
 Generally, the observed luminosities of the shell-type SNRs are consistent with the expected luminosity in 
the hadronic scenario. For RX~J0852.0-4622 (Vela Jr.), the distance estimates are controversial and therefore the
luminosity uncertainties can be as large as 2 orders of magnitude. See the references given in Table~\ref{table:SNR}
for more details.
Within this uncertainty on the distance of Vela Jr., the VHE observations are consistent
with the indirect evidence from X-ray observations for efficient shock acceleration of nuclei. 
A leptonic origin can not be excluded but is disfavored as it would lead to a comparably small efficiency for 
acceleration of nuclei which in turn would not produce non-linear modifications as are observed in X-rays.  
For Cassiopeia A, the average magnetic field is known to be larger than 80$~\mu$G \cite{1980MNRAS.191..855C}
and may be as large as 1~$\mu$G if magnetic field and particle energy density are in equipartition. This implies that the corresponding
particle number density and the expected inverse Compton flux is too small to explain the observed VHE emission which
effectively rules out a leptonic origin of the observed VHE emission in this particular case.\\
A second crucial test of the SNR origin of cosmic rays is the spectral shape and maximum energy of particles 
accelerated\footnote{which can not be inferred from observations in other wavelengths}. The observed VHE energy spectra
from RX~J1713.7-3946 \citep{2006A&A...449..223A} and RX~J0852.0-4622 \citep{2007ApJ...661..236A} indicate that 
the accelerated proton spectrum 
follows an $E_p^{-2}$ type power-law and cuts off at energies of roughly 100-200~TeV which is 
about one order of magnitude smaller than the energy of the ``knee'' in 
the cosmic ray all-particle spectrum at a few $10^{15}$~eV. This
discrepancy can be accommodated if these two SNRs are already too old and the shock has decelerated such 
that the maximum energy 
has dropped to the currently observed value \citep{2005A&A...429..755P} while the more energetic particles
have already escaped the remnant.\\
In this context, the young SNR Cassiopeia~A\footnote{and also Vela~Jr if the age is in the range of 200-300 yrs} 
and its VHE properties are crucial
to our understanding of the SNR contribution to cosmic rays.   
Cassiopeia A has been confirmed by the MAGIC and VERITAS collaborations to be a VHE-emitting source \citep{2007A&A...474..937A,2009AIPC..xxx..YYT} 
after the initial discovery with HEGRA \citep{2001A&A...370..112A}. The available spectra measured with HEGRA and MAGIC
  are in agreement with each other but require a rather soft energy spectrum with a photon index close to $2.5$. The total energy in 
protons $W_p\approx 2\times10^{49}$~ergs is only a small fraction of the kinetic energy of the blast wave and only 
a factor of 4-8 larger than the inferred energy in accelerated electrons \citep{2000A&A...355..211A}. 
  Given the young age of the source which is at the beginning of the Sedov phase,
  the small value of $W_p$ may still be re-reconciled with Cassiopeia A to be a typical SNR. 
It will be quite interesting to measure with better accuracy the
energy spectrum to higher energies in order to probe the maximum energy of accelerated protons which should be close to 
PeV energies. Furthermore,
as indicated in Table~\ref{table:SNR}, future detections (or stronger constraints) of VHE-emission from 
other historical SNRs like Tycho or Kepler will provide important clues on cosmic-ray acceleration in these objects.  

  \begin{table}
  \begin{minipage}{\textwidth}
  \caption{List of supernova-remnants (shell-type and mixed-morphology) observed in VHE gamma-rays \label{table:SNR}.}
  \begin{center}
   \begin{tabular}{lcccl}
    Name                      & distance & $L_\gamma$     & age & Type \\
    \hline	
                              & [kpc]    & [$10^{33}$ ergs/s]& [kyrs]   & \\
    \hline \hline
    Cas A                     & 3.4\footnote{\citet{1995ApJ...440..706R}, 
                                            $^b$\citet{2001A&A...370..112A}\citep{2007A&A...474..937A},
					    $^c$ \citet{2001AJ....122..297T},
					    $^d$ \citet{1997PASJ...49L...7K},
					    $^E$ \citet{2006A&A...449..223A},
					    $^e$ \citet{1997PASJ...49L...7K},
					    $^f$ \citet{2004A&A...427..199C},
					    $^G$ \citet{2007ApJ...661..236A},
					    $^g$ \citet{1999A&A...350..997A},
					    $^h$ \citet{2005ApJ...632..294B},
					    $^i$ \citet{2001ApJ...548..814S},
			                    $^j$ \citet{2006ApJ...648L..33V},	
				            $^k$ \citet{2007arXiv0709.4103H},
					    $^L$ \citet{2009AIPC..xxx..yyS},
				            $^l$ \citet{2002AJ....124.2145V},
					    $^m$ \citet{2008arXiv0801.3555H},
					    $^n$ \citet{1997ApJ...489..143C},
					    $^o$ \citet{2007ApJ...664L..87A},
				            $^p$ \citet{2008AJ....135..796L},
					    $^q$ \citet{1986MNRAS.219..427A},
				            $^r$ \citet{2001A&A...373..292A},
					    $^s$ \citet{1995A&A...299..193S},
				            $^t$ \citet{2003ApJ...585..324W},
					    $^u$ \citet{1999AJ....118..926R},
					    $^v$ \citet{2008A&A...488..219D},
					    $^w$ \citet{2007ApJ...668L.135R}
		 	}
                              & 1.4$^b$     
                              & 0.33$^c$
                              & Shell, II \\                                                                      
    RX~J1713.7-3946           & $\simeq 1^d$
			      & 5.7$^E$      
                              & $1.6^e$
			      & Shell, II/Ib$^f$\\
    Vela Jr                   & $0.2^g$  & 0.3$^G$ & 0.7$^g$ & Shell, II  \\
                              & $0.33^h$ & 0.6$^G$ & 0.66$^h$ & \\
                              & $1-2^i$  & 6-26$^G$& 4-19$^i$ & \\
    RCW 86                    & 2.8$^j$         
                              & 5.5$^k$ 
                              &1.8 (SN185?)$^j$
                              & Shell, ?\\
    SN~1006                   & 2.2$^t$  
			      & 1.4$^L$ 
			      & 1  
                              & Shell, Ia\\
    \hline
    W28                       &  1.9$^l$         & 0.5$^m$     &        $33^m$     & MM \\
    IC443                     &  1.5$^n$         & 0.4$^o$    &         $20^p$          & MM \\
    
   \hline
    \multicolumn{5}{c}{Without detection (historical SNe)} \\
   \hline
    Tycho	              & 2.2$^q$& $<0.5^r$ &0.4 & Shell, Ia \\
                              & 4.5$^s$& $<2^r$   &(SN1572)&           \\    
 Kepler                       & $4.8^u$ 
			      & $<2.4^v$  
			      & 0.4  
			      & Shell, Ia$^w$ \\ 
			      & $6.4^u$ 
			      & $<4.2^v$
			      & (SN1604) \\
\hline
   \end{tabular}
   \end{center}
   \end{minipage}
  \end{table}
  
\begin{figure}
\parbox{0.45\linewidth}{\includegraphics[width=\linewidth]{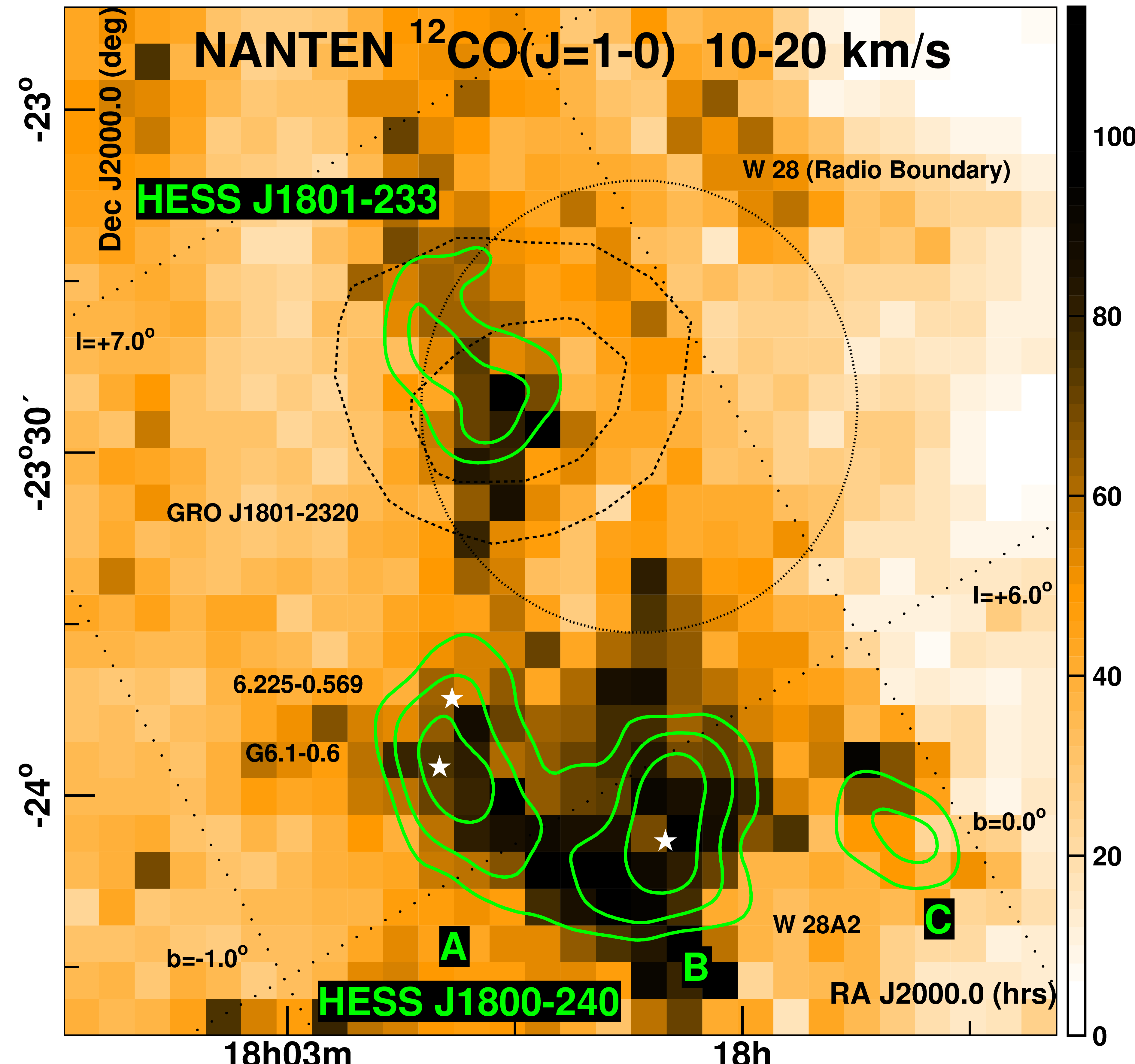}}
\parbox{0.55\linewidth}{\includegraphics[width=\linewidth]{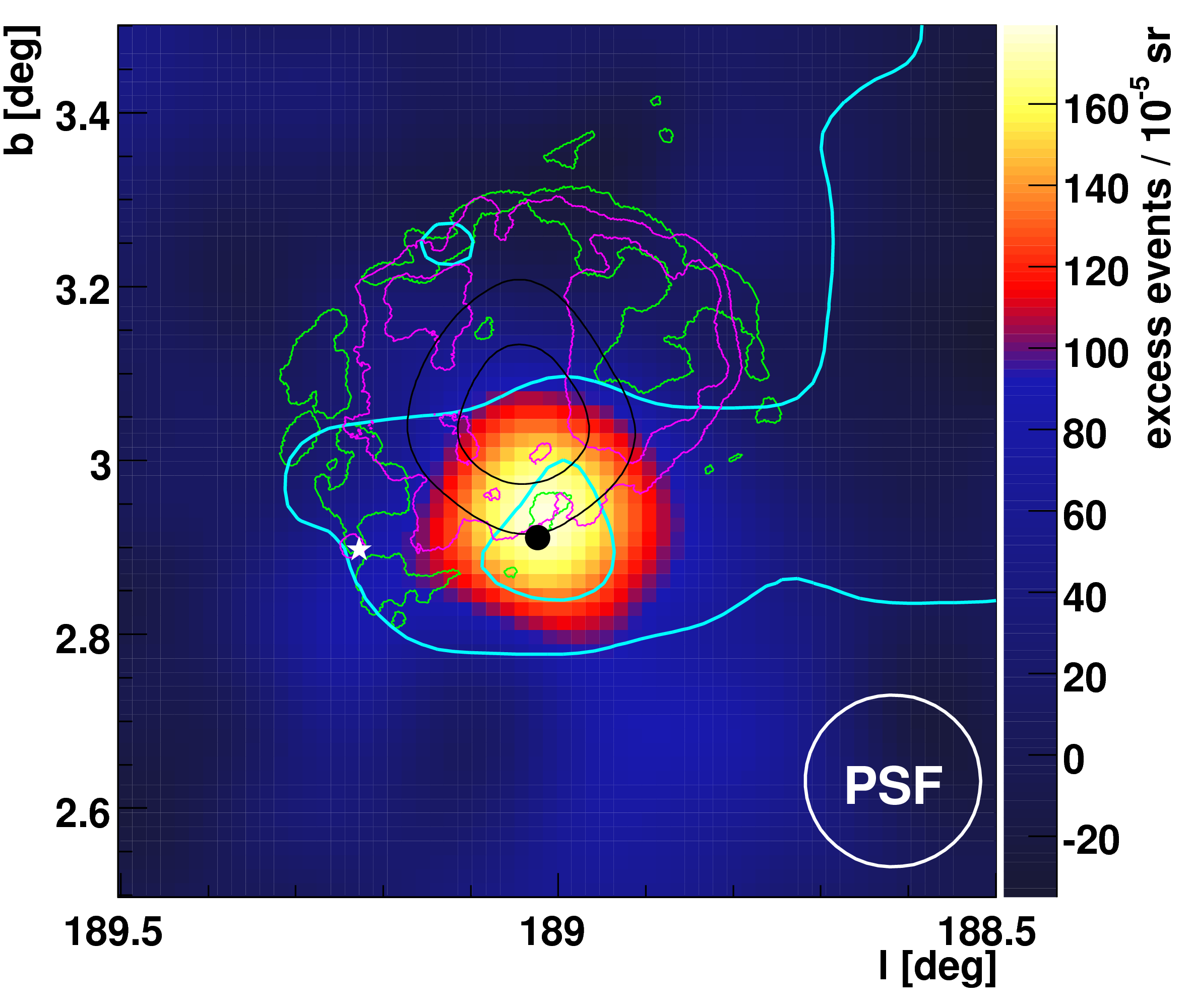}}
\caption{\label{image:MMR}Left panel (a): The integrated CO emission in the velocity range 10-20~km/s (color
scale) with significance contours from H.E.S.S. observations of W28 overlaid (green). The black contours 
follow the 95~\% and 99~\% confidence level regions for the unidentified EGRET source GRO~J1601-2320 \citep{2008arXiv0801.3555H}. Right panel (b): 
The color scale indicates the smoothed excess map obtained from MAGIC observations of IC~443. The molecular gas density (CO) is traced with the cyan colored contours, X-ray contours (purple) from ROSAT observations,
gamma-ray contours (EGRET) in black. See captions of Fig.~1 in \citet{2007ApJ...664L..87A} for references.
}
\end{figure}
 \subsection{Mixed morphology supernova remnants}

Besides young ($t\approx {\cal O}(1000)$~yrs) shell-type supernova-remnants, VHE-gamma-ray emission has also been detected from the direction
of mixed-morphology (MM) type supernova-remnants (SNRs) which are considered evolved systems. MM systems are characterized
by a shell-like emission observed \textit{only} in the radio-band with thermal, centrally-peaked X-ray emission 
predominantly from the interior of the SNR \citep[see e.g][]{1998ApJ...503L.167R}.  The mechanism responsible for heating the gas is not well-understood. 
It has been suggested that mixed morphology SNR have entered a post-Sedov stage of their evolution
which is characterized by an equal electron and ion temperature \citep{2005ApJ...631..935K}.
The fact that roughly 20~\% of the known Galactic SNR are of MM supports this scenario.\\
The observation of VHE gamma-rays from MM SNR like W28 \citep{2008arXiv0801.3555H} and IC443 \citep{2007ApJ...664L..87A} is -- independent of the 
underlying radiation mechanism -- direct evidence for particle acceleration to multi-TeV energies in these objects (see Fig.~\ref{image:MMR}). However, one would expect that the evolved and slow shock present in these objects is not strong enough to accelerate particles at the current stage of evolution. Furthermore, the observed Gamma-ray emission seems to coincide well with regions of dense molecular gas. In the north-eastern region of W28, 
some of the clouds must have already been interacting with the expanding blast wave. The interaction can be traced by maser emission in the shocked gas (see Fig.~\ref{image:MMR}a). 
The observation of VHE-emission from IC~443, another
MM type SNR, supports a similar scenario of an evolved SNR interacting with dense and cold
molecular gas (see Fig.~\ref{image:MMR}b). In this scenario, electrons can not produce the observed VHE-emission
because of the rapid cooling through inverse Compton and synchrotron emission, as
well as Bremsstrahlung in the dense gas. Typical cooling times of electrons at multi-TeV energies 
would be of the order of a few hundred years and by far too short to allow for a sizeable 
population of electrons to be present in an evolved SNR without ongoing acceleration. An alternative
and much more favorable interpretation for MM SNRs requires that accelerated nuclei have been partially confined
to the SNR and its environment and produce efficiently gamma-rays through $\pi^0$-decay in the dense target material of the molecular clouds. 
\citet{2007ApJ...665L.131G} have investigated the effect of  
cosmic rays that have left the accelerator and produce an observable VHE emission in a nearby cloud of molecular gas. 
%
 \subsection{Pulsar wind nebulae and composite supernova remnants} 
 While shell type SNR are considered to be the likely source of the Galactic cosmic rays, the 
blind survey of the Galactic plane \citep{2005Sci...307.1938A,2006ApJ...636..777A} performed with the H.E.S.S. telescopes has revealed a comparably small number
of shell-type SNR while VHE emission from pulsar wind nebula (PWN) systems has been observed more frequently \citep{2006ARA&A..44...17G}. 
In total, $20$ VHE sources (including candidate associations)
have been discovered which are very likely powered by isolated pulsars. \\ 
The population of ``TeV-Plerions'' \citep{2006A&A...451L..51H} includes
young objects like  Kes-75/PSR~J1846-0258 (spin-down age $\tau=723~$yrs) \citep{2007arXiv0710.2247H}, 
MSH 15-5\textit{2} \citep{2005A&A...435L..17A} driven
by PSR~J1513-5908 ($\tau=1.55$~kyrs) as well as 
evolved systems like the PWN driven by PSR~B1823-13 ($\tau=21.4$~kyrs) associated
with HESS~J1825-137 \citep{2005A&A...442L..25A}. Apparently, PWN
systems are active VHE sources for a few 10\,000 years thus exceeding the time that a SNR evolves through the Sedov-stage of shell type SNR during
which presumably VHE emission is most likely produced. The longer time during which PWNe are active particle accelerators explains the larger number
of PWN systems active at any given time in the Galaxy in comparison to shell-type SNRs.\\
 However, the contribution of PWN to the Galactic cosmic rays 
remains an open question as long as the nature of the particles accelerated in these systems is not clarified. Currently, for most
of the VHE emitting PWNe, a leptonic origin of the VHE emission can qualitatively explain the multiwavelength morphology and
energy spectra. As an example, HESS~J1825-137 has been studied intensively both in X-rays as well as in VHE-gamma-rays where
the source extension appears to be larger by a factor of $\approx 6$ than in the X-ray band. This larger size can be explained by a greater loss time of electrons radiating VHE gamma-rays via inverse Compton scattering than the loss time of electrons radiating synchrotron X-rays \citep{2006A&A...460..365A}. The energetic electrons injected by the
pulsar PSR~B1823-13 loose rapidly energy in the high magnetic field environment close to the pulsar where the X-ray nebula is observed \citep{2003ApJ...588..441G}. While the particles leave the high magnetic field environment, they presumably radiate only at wave-lengths longer than the X-ray band 
in the synchrotron channel while inverse Compton scattering off the cosmic microwave-background is the dominant energy loss 
process leading to observable, extended VHE emission.
In this
model, the slowly decaying spin-down power of the pulsar is accumulated in an expanding bubble containing 
the TeV-emitting electrons. This scenario is supported by softening of the observed VHE spectra with
increasing distance to the compact X-ray PWN system \citep{2006A&A...460..365A}. \\
The observed morphology of the PWN is often asymmetric and not centered on the pulsar. While
in some cases, the velocity of the pulsar can accommodate for the offset, other objects require an alternative 
explanation (e.g. Vela X): 
here, the asymmetric reverse shock of the SNR shell interacts with the relativistic PWN gas ''crushing'' and pushing it
off-centered \citep{2001ApJ...563..806B,2008A&A...478...17F}. \\
 A contribution to the observed VHE gamma-ray flux from nuclei accelerated by the pulsar is not excluded but seems to be less important than the contribution of electrons for most systems. \\
The Vela-X PWN system may
be an exception. Here, the spin down power of the pulsar is by far larger than the acceleration rate of electrons
required to drive the PWN system. In this sense, there is missing energy that can  be naturally explained by a nucleonic
component in the wind that drives the acceleration of electrons through resonant wave absorption \citep{1994ApJS...90..797A} and
produces the bulk of the observed VHE emission \citep{2006A&A...448L..43A}. The
VHE spectrum can be explained quite well in terms of over-all energetics 
(thus solving the problem of ''missing'' energy) 
as well as its shape \citep{2006A&A...451L..51H}.
 A crucial test will be the detection of high energy neutrinos from this object that may 
be the brightest steady neutrino source in the sky \citep{2007ApJ...656..870K, 2007ApJ...661.1348K}. \\
Finally, composite systems where a plerionic, non-thermal X-ray component is observed to be embedded in a radio-emitting shell
like G0.9+0.1  have been found to emit VHE gamma-rays \citep{2005A&A...432L..25A}. In addition to known composite SNRs, VHE observations have helped to discover and identify new systems: X-ray observations of HESS~J1813-178 \citep{2005Sci...307.1938A} carried out with XMM-Newton 
 have revealed a previously unknown extended X-ray source inside the
radio-shell of SNR G12.82-0.02 \citep{2007A&A...470..249F} putting this object in the category of composite SNR systems. Most likely, the VHE-emission is produced in the central PWN systems while the radio-shell is VHE-dim. However, it should be noted, that the spatial extension of these two systems is barely
resolved at VHE energies.

\subsection{X-ray binary systems}
Before the imaging air Cherenkov technique was widely established, several groups had claimed 
a number of X-ray binary sources to be periodic (e.g Her X-1, Cyg X-3). 
Cyg X-3 had been claimed to be a   transient source
of energetic particles up to PeV energies \citep{1983ApJ...268L..17S}, 
for a review of early claims of detections see \citet{1992SSRv...59..315W}. With the notable
exception of a possible burst observed from the direction of GRS~1915+105 \citep{1998NuPhS..60..193A}, 
none of the XRB systems had been confirmed until recently
to be VHE gamma-ray emitters. 
Finally 4 XRB systems listed in Table~\ref{table:XRBs} have been detected to be VHE gamma-ray sources. 
All of these objects show variability in the VHE band either linked to the orbital motion as for 
LS~I~+61~303 \citep{2006Sci...312.1771A,2007arXiv0709.3661M}, 
LS~5039 \citep{2006A&A...460..743A}, and PSR~B1259-63 \citep{2005A&A...442....1A}, 
or transient activity as for Cyg X-1 \citep{2007ApJ...665L..51A}.\\
Given that these objects are currently the only variable Galactic sources of VHE gamma-rays, 
the interpretation of their properties is quite different
from the other source types. 
The orbital parameters as well as the physical conditions in these systems are reasonably 
well-known and allow for a detailed modelling of the processes leading to acceleration and VHE  
emission. In this
sense, one can consider the systems as ''laboratories''.

\begin{table}
\caption{X-ray binary systems detected with imaging air Cherenkov telescopes.\label{table:XRBs}.}
\begin{minipage}{\textwidth}
\begin{tabular}{llcccl}
\hline
Identifier           & Variability             & D & $L$\footnote{Integrated from 1 to 10 TeV} & System\footnote{BH: Black hole candidate, NS: Neutron star, PSR: Pulsar} \\
                     &                         & [kpc]    & [$10^{33}$~ergs/s]  & \\
\hline \hline
PSR~B1259-63 & $P=3.4$~a             &1.5	  &  1.6        &PSR(47~ms)/B2e \\
LS~5039              & $P=3.9$~d            &2.5    &  8.7        &BH or NS/O6.5V    \\
LS~I~+61~303         & $P=26.5$~d           &2	  &  2.5        &NS/B0Ve   \\
Cyg X-1              & $T_\mathrm{var}\approx 80$~min&2.2    &  1.6        & BH/O9.7 Iab \\
\hline

\end{tabular}
\end{minipage}
\end{table}

 The pulsar PSR~B1259-63 ($P=48$~ms) is in a highly eccentric orbit ($P_\mathrm{orbit}=3.4$~yrs)
 around the Be-type companion star SS2883. VHE emission
is predominantly produced around the periastron where the effects of adiabatic and 
radiative cooling due to inverse Compton as well as synchrotron emission are 
strongest \citep{2005A&A...442....1A}.
 The IC component had been predicted successfully prior to the detection \citep{1999APh....10...31K},
albeit the observed temporal behavior of the source as it passed through the  periastron in 2004 
was quite different from the expectation \citep{2005A&A...442....1A}. Refined models including a 
more detailed treatment of inverse Compton scattering in energetic and anisotropic radiation fields 
as well as adiabatic energy losses can reproduce the observed light curve
and provide some crucial predictions to be probed with new observations \citep{2007MNRAS.380..320K}.  
Alternative models invoking the presence of a nucleonic component that leads to VHE emission
through the interaction with the disk-outflow of the Be-star have been proposed as well \citep{2004ApJ...607..949K,2007Ap&SS.309..253N}. \\
Unfortunately, the periastron passage in 2007 took place during the time when PSR B1259+63 
culminated during the day time. A crucial 
observation will be only possible during the  periastron passage in 2011 when the system will
 be visible for ground based VHE instruments as well as for Fermi.   

\begin{figure}
\includegraphics[width=\linewidth]{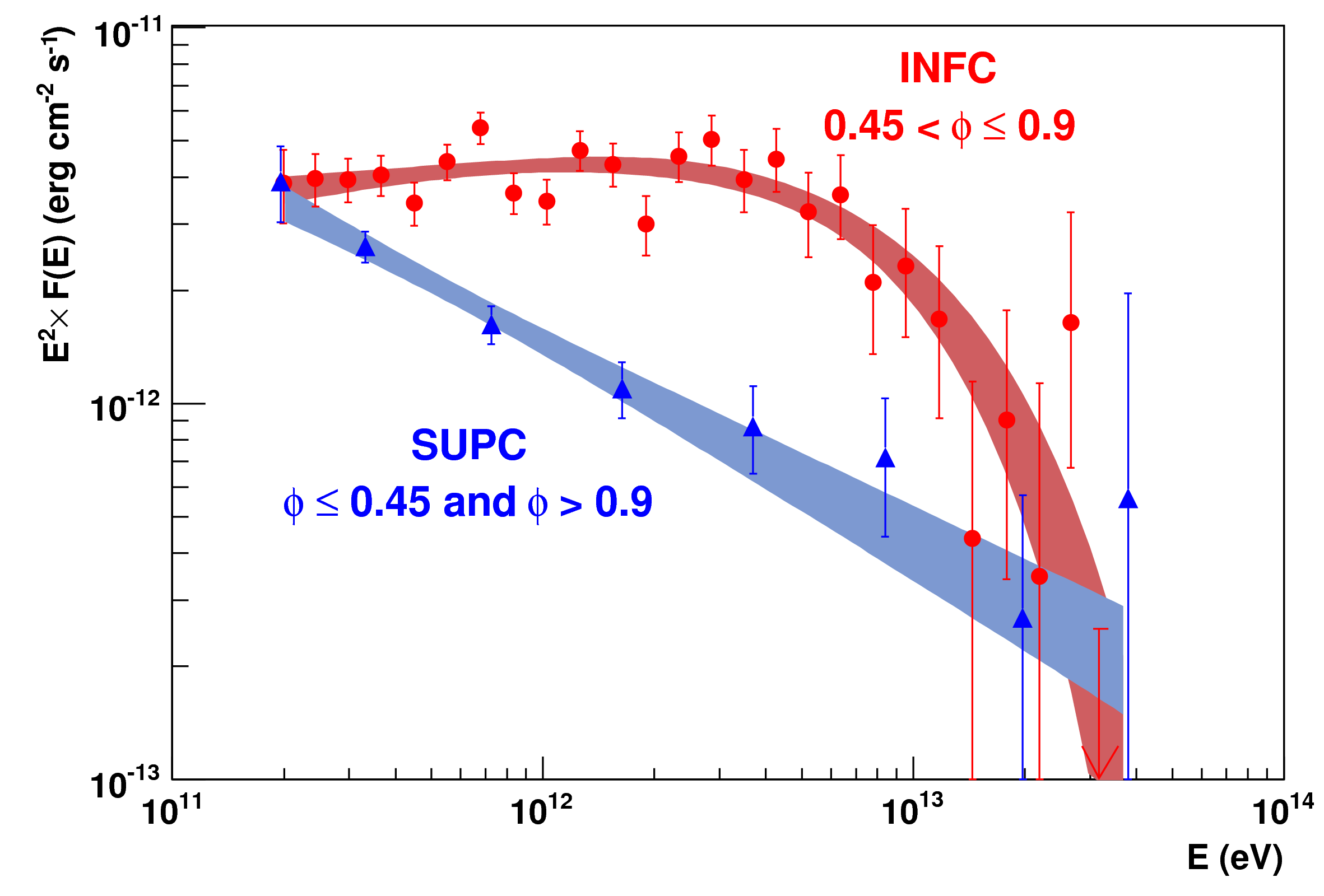}
\caption{The VHE energy spectrum of LS~5039 for two distinct orbital phases. The \textit{INFC} corresponds
to the inferior conjunction, while \textit{SUPC} is the phase interval at the superior conjunction with 
the periastron passage \citep{2006A&A...460..743A}. \label{ls5039:spec}}
\end{figure}

The two systems LS 5039 as well as LS~I~+61~303 have shorter orbital periods of 3.9 and 26.5 days, 
respectively.  Initially, both objects have been considered to be micro-quasars with a 
steady radio-jet feature.  There is an ongoing debate whether this picture may  have to be
modified. Radio observations of LS~I+61~303 indicate that the orientation of the ''jet'' 
changes during the orbit \citep{2006smqw.confE..52D} which is more in line with the cometary tail 
predicted by \citet{2006A&A...456..801D}. The cometary tail is  a consequence
of the interaction of a pulsar wind with the stellar outflow. For LS~5039, the current high resolution
radio observations also indicate a change in the orientation of the jet-like structure 
\citep{2008A&A...481...17R}. However, since the observations do not cover a full orbital period, 
the results are not conclusive yet. \\
Given the uncertainty of the nature of the compact object, 
the VHE emission from LS~5039 has been interpreted in 
the context of a micro-quasar model \citep{2006A&A...447..263B}
as well as in an interacting PWN model \citep{2006A&A...456..801D}. \\
A crucial consideration is the absorption of VHE photons due to pair-production in the photon field of the stellar companion. If the line-of-sight towards the VHE emission region passes within roughly $10^{14}$~cm of the companion star of LS~5039,
absorption will produce visible effects \citep{2006A&A...451....9D} varying with the phase of the orbit. 
The absorption effect is energy dependent and should lead to a pronounced variability at energies between 200-400~GeV. 
However, the observations indicate that the flux at 200-400~GeV remains almost unchanged along the orbit while
the modulation is most pronounced at TeV-energies (see Fig.~\ref{ls5039:spec}).\\
Secondary particle production in cascades can in principle reproduce the observed variability \citep{2007A&A...464..259B} as well
as models where the emission takes place at a larger distance to the stellar companion \citep{2008MNRAS.383..467K}. 

\subsection{Star-forming regions}
\begin{figure}
\parbox{0.5\linewidth}{
\includegraphics[width=\linewidth]{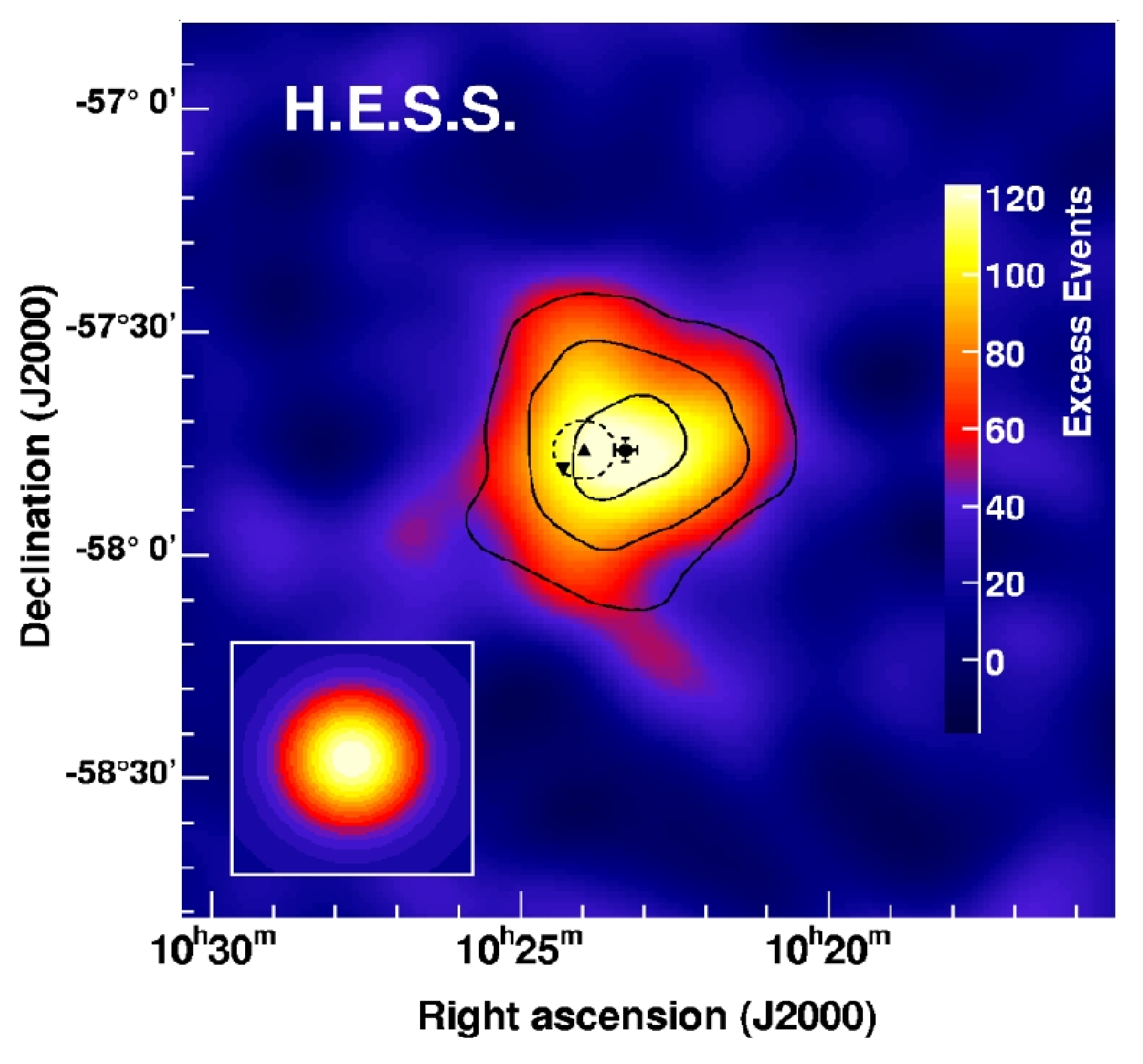}
}
\parbox{0.45\linewidth}{
\caption{Westerlund 2 \citep{2007A&A...467.1075A}: 
The color scale indicates the smoothed VHE excess map (see the bottom left inlaid figure for the 
point-spread function), the contours are the significance levels starting at $5~\sigma$ 
and increasing by $2$.  The filled circle and error bars mark the position and uncertainty of 
the VHE source ($1~\sigma$). The dashed circle marks the position and extension of Westerlund 2, 
the upright triangle is at the position of the Wolf-Rayet binary system 
WR~20a and the downward triangle is at the position of WR~20.\label{fig:wd2}}
}
\end{figure}

So far, the objects under scrutiny are related to the final stages of stellar evolution and their 
remnants. However, gamma-rays are also produced in molecular clouds which are the cradles of star 
formation and possibly during the final mega-year of stellar evolution where massive stars 
($>15~M_\odot$) drive powerful winds. 
The wind-phase of stellar evolution is characterized
by mass loss rates of up to $10^{-5}~M_\odot$/year and terminal velocities of a few 1000 km/s. 
These early-type stars are usually born in open stellar associations in the spiral 
arms of the Galactic disk. The massive O stars in these associations often evolve into 
the Wolf-Rayet phase where strong stellar winds produce a hot tenuous cavity in the interstellar 
medium. In such systems up to $10^{39}$~ergs/s are dissipated in the form of kinetic energy 
from stellar winds, even before member stars
evolve into supernovae. Provided that shocks will form either
in the termination region of cumulative winds \citep{2006A&A...448..613D} or in wind-wind interaction 
regions \citep{2006ApJ...644.1118R, 2007A&ARv..14..171D}, some of this power can be converted into 
cosmic-ray acceleration.  \\
The detection of VHE-emission from Westerlund-2 is a crucial step towards establishing the 
r\^ole of stellar driven cosmic ray acceleration (see Fig.~\ref{fig:wd2} from \citet{2007A&A...467.1075A}). 
Westerlund-2 is a young star forming region with roughly 20 O-type stars and the Wolf-Rayet systems 
WR 20a\&b \citep{2007A&A...463..981R}. The total power released in the stellar winds is estimated to 
be $5\times 10^{37}$~ergs/s. The observed VHE emission can be easily produced if roughly 
1~\% of the kinetic energy is converted into the acceleration of electrons 
\citep{2007A&A...474..689M,2007arXiv0710.3418R}.
\\ 
Prior to the discovery of VHE emission from Westerlund-2, the first unidentified VHE source 
TeV~J2032+4130 had been associated with the remarkably powerful OB association in 
the Cygnus arm. The HEGRA discovery of TeV~J2032+4130 \citep{2002A&A...393L..37A,2005A&A...431..197A} 
has been recently confirmed by MAGIC \citep{2008arXiv0801.2391A} as well as by the MILAGRO and 
VERITAS groups \citep{2007ApJ...658L..33A, 2007ApJ...664L..91A,2004A&A...423..415L,2007ApJ...658.1062K}. 
Remarkably, observations with the MILAGRO detector show an extended 
($\approx 3^\circ$) source  of $>10$~TeV photons centered
on TeV~J2032+4130 \citep{2007ApJ...658L..33A}. This source has
received considerable multi-wavelength coverage including
deep radio observations which show indications for
extended radio emission coinciding with the VHE source \citep{2007ApJ...654L.135P}.  Recent 
X-ray observations with XMM-Newton have revealed faint non-thermal extended X-ray emission co-located 
with the VHE-source \citep{2007A&A...469L..17H}.\\   
 Future observations of this region with the VERITAS telescope array will be very helpful to 
disentangle possible multiple-sources and to study the source size at different energies.  \\
Besides Westerlund-2 and Cyg OB2,  the MILAGRO source MGRO~J2019+37 \citep{2007ApJ...658L..33A} 
has been  associated with the young stellar association
Berk-87 including the Wolf-Rayet object WR~142 \citep{2007MNRAS.382..367B}. \\ 
A systematic search for VHE emission with the HEGRA telescopes from young open clusters has so far produced only upper limits \citep{2006A&A...454..775A}.  

\subsection{Unidentified Galactic sources}
\begin{figure}
\parbox{0.4\linewidth}{
\includegraphics[width=\linewidth]{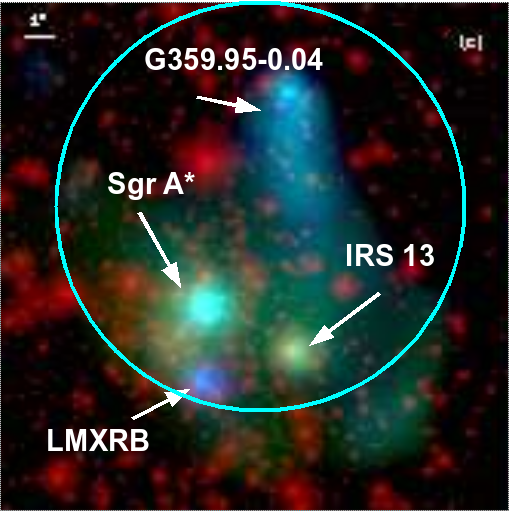}
}
\parbox{0.58\linewidth}{
\includegraphics[width=\linewidth]{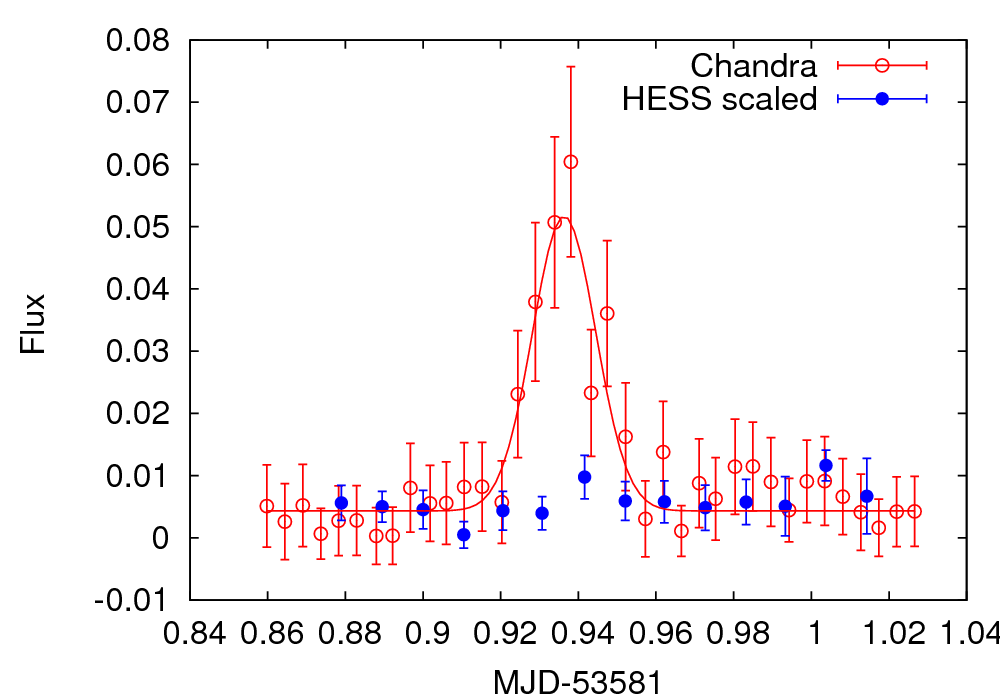}
}
\caption{\textbf{Left panel:} Chandra X-ray and SINFONI near-IR composite from \citet{2006MNRAS.367..937W} with additional 
annotations. The circular region corresponds to the combined systematic and statistical uncertainty for
the HESS-source associated with the Galactic center.\label{fig:center}
 \textbf{Right panel:} X-ray and VHE-gamma-ray light curve from simultaneous observations of Sgr A*. The H.E.S.S. light curve has
been scaled to match the Chandra light curve in units of counts per second. Note the comparably small error bars
of the H.E.S.S. observations - a flare of similar magnitude as observed in the X-ray band would have been clearly detected.}
\end{figure}

 The sources discussed in the previous section have been identified or are associated with known objects. 
However, a number of objects remain to be identified. A total of 10--20 objects have no clear or 
even multiple counterparts. Among these objects, 
the Galactic center source and the point-like source in the Monoceros region are highlighted in the 
following. For a recent summary of unidentified sources, see e.g. 
\citet{2007Ap&SS.309...11F,2008A&A...477..353A}.
\paragraph{Galactic center region.}
 The central region of our Galaxy contains a super-massive black hole with a mass of 
$3\times 10^6~M_\odot$ \citep{2002Natur.419..694S,2005ApJ...620..744G}. 
Given the faintness of this object, the accretion rate must be  $<10^{-8}$ of the Eddington-rate. The 
mm-, IR-, and X-ray sources associated with Sgr A* have been observed to show variability 
including flares and outbursts with a time-scale of hours \citep{2003Natur.425..934G}. The 
shortness of the flares constrains the emission region to be compact with a spatial extension 
$r<t_\mathrm{var}\cdot c\approx  10^{14}\mathrm{~cm}~(t_{var}/hr)\approx 100 r_G (M/3\cdot 10^6 M_\odot)^{-1} t_{var}/\mathrm{hr}$. Observations 
with the CANGAROO III telescopes as well as at large zenith angles with the Whipple 10m telescope 
revealed a VHE gamma-ray source co-located with the Galactic center 
\citep{2004ApJ...606L.115T,2004ApJ...608L..97K}. The positional accuracy as well as 
spectral measurements have been considerably improved 
with the observations of the H.E.S.S.-telescopes \citep{2004A&A...425L..13A}. The MAGIC telescope
has confirmed the earlier findings in independent observations \citep{2006ApJ...638L.101A}. \\
The detection of a point-like source with a hard power-law spectrum was shortly after
the discovery discussed in the context of emission scenarios near to the super-massive black hole 
(at distances larger than $\approx 10~r_G$ in order to avoid
internal absorption) from e.g. ultra-high energy protons emitting  synchrotron and curvature 
radiation in the high magnetic fields that are believed to be
threaded in the accretion disk \citep{2005ApJ...619..306A} or $\pi^0$-decay 
from energetic protons \citep{2007ApJ...657L..13B, 2006ApJ...647.1099L}. At 
a larger distance from Sgr A*, acceleration in SNRs \citep{2007JPhG...34.1813E} including 
Sgr A East \citep{2005ApJ...622..892C} as well as acceleration in stellar winds have been 
proposed \citep{2005ApJ...635L..45Q}.\\
The initial data were consistent with a Dark Matter annihilation scenario albeit requiring 
uncomfortably large masses beyond 20~TeV for the annihilating particles \citep{2005PhLB..607..225H}.
\\
 With more data, the H.E.S.S. experiment has constrained the 
energy spectrum to extend beyond 10~TeV  and to deviate from a WIMP-annihilation scenario 
\citep{2006PhRvL..97v1102A}
\footnote{see however \citet{2008JHEP...01...49B} for a discussion on 
the effects of internal Bremsstrahlung photons which are important for heavy WIMP annihilation in 
order to reduce the helicity suppression for the annihilation channel}. 
With the increased exposure, faint spatially extended VHE emission along the Galactic ridge 
was detected \citep{2006Natur.439..695A}.
The accuracy of reconstruction the position of the VHE-emitting point source has been improved to 
the level of the systematic pointing  uncertainty of $\approx 6$ arc~seconds 
\citep{2007arXiv0709.3729V} and excludes Sgr A East. 
The error box for the VHE source still encompasses at least three possible candidates for 
VHE emission (see Fig.~\ref{fig:center}a):
Sgr A*, a low-mass  X-ray binary system (LMXRB) \citep{2005ApJ...633..228M}, the 
stellar system IRS~13 \citep{2003ApJ...591..891B}, and the PWN G359.95-0.04 \citep{2006MNRAS.367..937W}. 
The PWN system has been argued to be a possible 
candidate to explain the observed VHE emission \citep{2007ApJ...657..302H}.\\
An association with a variable source like Sgr A* would be supported if simultaneous multi-wavelength 
(MWL) observations of Sgr A* during an outburst would show a correlation between e.g. 
X-ray and gamma-ray flux. 
During a dedicated MWL observation of Sgr A* in summer 2005 with H.E.S.S. and the Chandra
X-ray telescope, a typical X-ray flare was detected with the Chandra instrument 
\citep{2007arXiv0710.1537H}. The simultaneously taken H.E.S.S.-data show however no variability even
though the observed rate of VHE photons is sufficiently large to detect a flare of similar 
strength as seen in the Chandra data (see Fig.~\ref{fig:center}b). The 
absence of correlation would disfavor at least a common origin of the X-ray and VHE emission from Sgr~A*. \\
An association of the VHE source with the LMXRB could in principle be demonstrated by correlated 
activity in different wavelength-bands. However, so far no such observations are available.  

At this point it is not possible to discern between the different source candidates nor exclude 
a dark matter origin. 

\begin{figure}
\mbox{\parbox{0.5\linewidth}{\includegraphics[width=\linewidth]{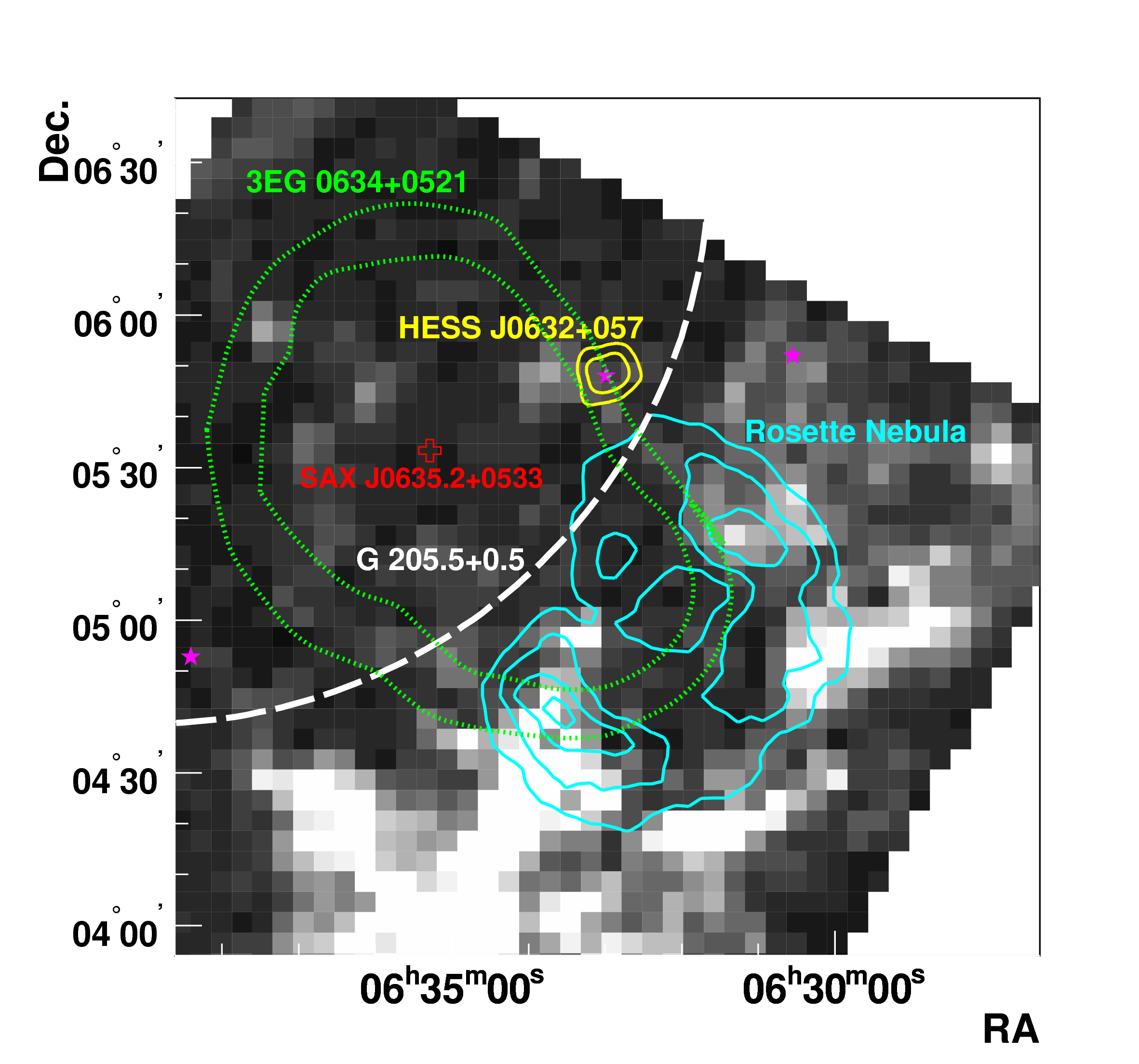}}
\parbox{0.5\linewidth}{
\caption{\label{fig:mono} Monoceros source \citep{2007A&A...469L...1A}: The grey-scale illustrates the velocity-integrated (0-30~km~s$^{-1}$) $^{12}$CO
emission as measured with the NANTEN telescope \citep{2004ASPC..317...59M}.
 The yellow-contours trace the 4 and 6~$\sigma$ 
level of detection with the H.E.S.S. telescopes. The cyan contours indicate the 8.35~GHz radio measurements
from \citet{2000AJ....119.2801L} while the green contours mark the 95~\% and 99~\% confidence regions for the position of the
EGRET source 3EG~0634+0521. SAX~J0635.2+0533 is a pulsar binary system. The positions of Be-stars are 
indicated with pink stars.} 
}}
\end{figure}

\paragraph{Monoceros source.}
The newly discovered population of unidentified VHE sources in the Galactic plane form a quite 
homogeneous sample of objects with respect to their spatial extension and energy spectra. It 
should be emphasized however, that the similarity of the VHE characteristics does not necessarily
imply that the underlying sources are similar. Certainly, observational biases are introduced by 
e.g. the limited sensitivity for detecting sources with low surface brightness.  \\
Among the unidentified sources, there is - besides the Galactic center - only one more 
unresolved object: HESS~J0632+058 which is close to the
rim of the Monoceros Loop SNR \citep{2007A&A...469L...1A}. The radial extension of this object 
is constrained to be smaller than $2'$ (95\% c.l.). 
No strong indications for variability are observed.  
The energy spectrum of the source is well-described by a power-law ($dN/dE\propto E^{-2.5\pm 0.3}$)
which is at the soft end of the distribution of energy spectra observed from unidentified VHE-sources. \\
For an unresolved source ($r<2'$), the search for a counterpart is simpler than for extended objects.
 In Fig.~\ref{fig:mono}, the position of HESS~J0632+058 is combined with multi-wavelength observations. 
Only a few candidate objects
coincide spatially with the VHE source including an unidentified EGRET source 3EG~J0634+0521 \citep{1999ApJS..123...79H}, an unidentified X-ray source 1RXS~J063258.3+054857 \citep{2000yCat.9029....0V}, and
a B0pe type star MWC~148 which may be a binary system similar to PSR~B1259-63. 
Alternatively, it could be 
a new type of VHE source associated with an isolated stellar system.
\subsection{Extra-galactic sources}
 The blind survey of the Galactic plane has revealed an interesting and previously unknown population of VHE-sources. 
Unfortunately, a survey with similar sensitivity of the extra-galactic sky has to be postponed until wide-field-of-view Cherenkov 
telescopes or the next generation of Water-Cherenkov-experiments like HAWC will come online (see also next section). 
Currently, the observations have focussed
on Blazars ("BL Lac type Quasars"). \\
 Historically, the discovery of VHE emission from the Blazars Mkn~421 \citep{1992Natur.358..477P} 
and Mkn~501 \citep{1996ApJ...456L..83Q} have opened the
field of TeV-Blazar observations. These objects are in general characterized by a featureless optical spectrum,
non-thermal X-ray emission, and violent variability in the optical wavelength-band as well as in X-rays. The spectral energy distribution 
is dominated by two broad maxima - one in the optical-to-X-rays and one in the gamma-ray-to-VHE-gamma-ray domain. The low-energy maximum is usually
assumed to be produced by synchrotron emitting electrons while the high energy bump would be due to inverse-Compton scattering. \\
TeV-Blazars can be considered ``clean'' systems where the main source of seed photons are actually the synchrotron photons emitted by 
the electrons themselves, commonly called Synchrotron-Self-Compton (SSC)
mechanism. Within the unified scheme, Blazars are believed to be Fanaroff-Riley
Type I galaxies with a jet axis pointing close to the line-of-sight. 
\\ 
In a simple one-zone-SSC model, the VHE luminosity is closely linked to the
synchrotron luminosity which in turn is related to the particle density
and provides seed photons for inverse Compton scattering.
A simple but very successful selection scheme of possible VHE
emitting candidates is based on the X-ray or radio flux of Blazars
\citep{2002A&A...384...56C}.  Most of the Blazars that have been predicted to be
detectable according to this scheme have been actually discovered by the H.E.S.S.
and MAGIC collaboration. 
\\ 
With the growing number of known Blazars at different red-shifts (see
Table~\ref{blazartable} for a list of extra-galactic VHE-gamma-ray sources), a
number of issues and questions can be addressed: 
\begin{itemize} 
\item Is a simple one-zone SSC model sufficient to explain all (multi-wavelength) data
including correlated variability patterns or is a more complicated model
required (stratified jets, multiple zones model)?  
\item Where does the acceleration take place?  
\item What is the underlying acceleration mechanism in
Blazars and what is the composition of the jet? This is a fundamental question
related to the jet physics and the energetics of AGN.  
\item What is the duty cycle of Blazars?  
\item How strong is the beaming? In the relativistic outflow,
a co-moving isotropic emission is strongly beamed in the forward direction. With
increasing bulk Lorentz factor $\Gamma$, the opening angle gets narrower
$\propto \Gamma^{-1}$.  The higher the beaming, the smaller the likelihood that
we can observe the emission. In consequence, a high beaming factor can lead to
dramatically increased number of Blazars that are pointing \textit{not} in our
direction.  
\item Is there internal absorption (due to photon-photon scattering
with pair-production)?  
\item How strong is the absorption on the extra-galactic
background light (EBL)? 
 \end{itemize}

In the following, I will highlight observations from Blazars as well as from
non-Blazar VHE sources including M87 and the flat-spectrum-radio quasar 3C279. 

\subsection{Blazars}
\begin{figure}[ht!]
\includegraphics[width=\linewidth]{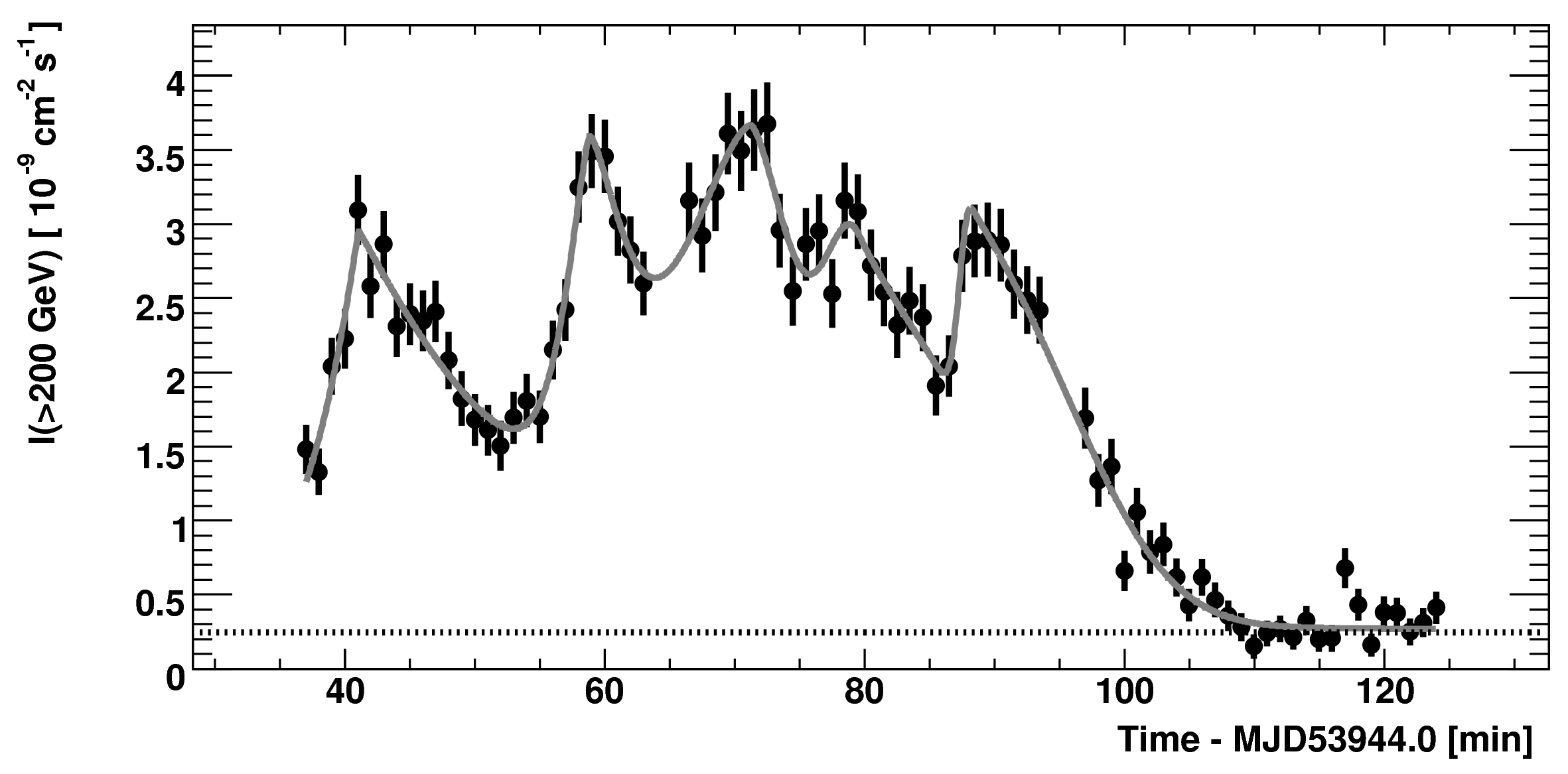}
\caption{\label{flares:2155} 
The light curve from PKS~2155-304 in the night July-28 2006 as measured with
H.E.S.S. above 200 GeV.  The data are binned in 1-minute intervals. For
comparison, the horizontal dotted line represents I($>$200 GeV) observed from
the Crab Nebula. The curve is a fit of a superposition of five individual bursts
and a constant flux.  See \citet{2007ApJ...664L..71A} for additional details.}
\end{figure}

 A very spectacular feature of VHE emission from Blazars is the 
short time-scale variability.  The most impressive examples of short flares have
been observed in 2006 from the nearby object PKS~2155-305 (red-shift $z=0.116$).
This object was the first Blazar discovered in the southern hemisphere by the
Durham group using their Mark VI telescope in Australia
\citep{1999ApJ...513..161C}.  The H.E.S.S. telescopes observed their ``first
gamma-light'' from this source already in 2002 when the first of the four
H.E.S.S. telescopes went online \citep{2007icrc..punch}.  Since PKS~2155-305 is
a remarkably bright object, it can be easily detected in quiescent state with
the system of four H.E.S.S. telescopes. Detailed measurements of the energy
spectrum show despite widely different flux states only marginal deviations from a
soft power-law with photon index $\approx 3.3$ \citep{2005A&A...430..865A}.
PKS~2155-305 has been the target of a number of simultaneous
multi-wavelength-campaigns together with the RXTE X-ray satellite as well as in
conjunction with the Chandra, XMM-Newton, and the SWIFT X-ray telescopes
\citep{2005A&A...442..895A}. 
During summer 2006, the source went into an unprecedented high state reaching
flux levels more than a factor of 100 higher than the quiescent flux
\citep{2007ApJ...664L..71A}. During the peak flares, the rate of detected VHE
gamma-rays reached values exceeding 1 Hz which marks the highest rate ever
detected at VHE energies.  During the night with the highest flux, a sequence of
flares can be discerned with rise and decay timescales of minutes (see
Fig.~\ref{flares:2155}). This immediately constrains the emission region to be
very compact and to be moving with a large bulk Lorentz factor to overcome
pair-opacity and to relax the size constraint \citep{2008MNRAS.384L..19B}. 
Under the assumption that the variable emission is produced close to the super-massive
black hole, two interesting conclusions have been pointed out recently:
(a) The variability time scale constrains the mass of the black hole to
be of the order of $10^7~M_\odot$ roughly 2 orders of magnitude smaller than
the general correlation of the mass with luminosity of the host galaxy suggests
\citep{1998MNRAS.293..239C,2008arXiv0806.2545N,2003A&A...399..869B}.
(b) The apparent periodicity of the subflares observed could potentially
provide a measure of the angular momentum of the black hole \citep{2008arXiv0806.2545N}.
 \subsection{Other active galactic nuclei}
 Besides the Blazars, only two non-Blazar extra-galactic objects have been
discovered so far: M~87 and 3C~279\footnote{A signal from Centaurus A at energies
above 300~GeV had been claimed on the basis of non-imaging observations of air
Cherenkov light at the level of 4.5~$\sigma$ \citep{1975ApJ...197L...9G}. This
claim awaits confirmation.} 

\begin{table}
\begin{minipage}{\textwidth}
  
\caption{List of extragalactic VHE-sources sorted by red-shift. Whenever available, different 
source activity states are characterized \label{blazartable}.}
\begin{center}
\small
\begin{tabular}{lcccccl}
\hline
Source &
Type   & 
$z$  & 
$d_L$\footnote{Assuming $\Omega_m=0.27$, $\Omega_\Lambda=0.73$, $H_0=73$km~s$^{-1}$Mpc$^{-1}$}     &
$\log_{10}\left({L_\gamma}\right)$\footnote{Isotropic luminosity $L_\gamma$ in ergs~s$^{-1}$ integrated between 0.1 and 1~TeV} &
$\Gamma_\mathrm{measured}$ &
Ref. \\
\hline
M87     	& FR I & 0.0044     &23   &41.1 & $2.6\pm0.4$ &
\footnote{\citet{2006Sci...314.1424A},
$^d$\citet{2002A&A...393...89A},
$^e$\citet{2007ApJ...669..862A},
$^f$\citet{2007ApJ...662..892A},
$^g$\citet{2005ApJ...634..947S},
$^h$\citet{2006ApJ...648L.105A},
$^i$\citet{2006ApJ...639..761A},
$^j$\citet{2003A&A...406L...9A},
$^k$\citet{2007ApJ...666L..17A},
$^l$\citet{2007arXiv0710.4057T},
$^m$\citet{2005A&A...436L..17A},
$^n$\citet{2008arXiv0802.4021H},
$^N$\citet{2008arXiv0808.0889V},
$^o$\citet{2005A&A...430..865A},
$^p$\citet{2007ApJ...664L..71A},
$^q$\citet{2002A&A...391L..25D},
$^Q$\citet{2008ATel.1415....1S},
$^r$\citet{2007A&A...475L...9A},
$^s$\citet{2006A&A...455..461A},
$^t$\citet{2006ApJ...642L.119A},
$^u$\citet{2007A&A...473L..25A},
$^v$\citet{2007A&A...470..475A},
$^w$\citet{2007ApJ...667L..21A},
$^x$\citet{2008A&A...477..481A},
$^X$\citet{2008A&A...487L..29N},
$^Y$\citet{2008ATel.1500....1T},
$^y$\citet{2008Sci...320.1752A}} \\
                &      &            &     &41.6 & $2.2\pm0.2$ & $^c$\\  
Mkn~421 	& HBL  & 0.030      &130  &44.8 & $3.0\pm0.2$  & $^d$\\
                &      &            &     &45.3 & $2.06\pm0.03$& $^d$\\
Mkn~501 	& HBL  & 0.034      &142  &44.3 & $2.45\pm0.07$& $^e$\\
                &      &            &     &45.2 & $2.09\pm0.03$& $^e$\\ 	
1ES~2344+514     & HBL  & 0.044     &183  &43.9 & $2.9\pm0.2$  & $^f$\\
                 &      &           &     &45.2 & $2.5\pm0.2$  & $^g$\\
Mkn~180 	& HBL  & 0.045      &194  &44.0 & $3.3\pm0.7$  & $^h$ \\
1ES~1959+650 	& HBL  & 0.047      &198  &44.2 & $2.7\pm0.1$  & $^i$ \\
                &      &            &     &44.9 & $1.8\pm0.2$  & $^j$ \\
BL Lacert\ae    & LBL  & 0.069      &293  &44.0   &$3.6\pm0.5$ & $^k$\\
PKS~0548-322    & HBL  & 0.069      &300  &43.6   &$2.8\pm0.3$ & $^l$\\
PKS~2005-489    & HBL  & 0.071      &306  &44.2   &$4.0\pm0.4$ & $^m$\\
RGB~J0152+017   & HBL  & 0.080      &345  &44.0   &$3.0\pm0.4$& $^n$\\
W Comae         & IBL  & 0.102      &452  &44.9   &$3.8\pm0.4$& $^N$\\ 
PKS~2155-304    & HBL  & 0.116      &515  &44.1   &$3.4\pm0.1$ & $^o$\\
                &      &            &     &46.8   &$2.7\pm0.1$ &$^p$  \\
                &      &\multicolumn{3}{r}{Break at 0.43~TeV}  &$3.5\pm0.1$ & $^p$\\
H~1426+428      & HBL  & 0.129      &585  &45.9   &$3.7\pm0.4$ & $^q$\\
1ES~0806+524    & HBL  & 0.138      &626  &43.8   & assumed 2.6& $^Q$  \\
1ES~0229+200    & HBL  & 0.139      &633  &44.3   &$2.5\pm0.2$ & $^r$\\
H~2356-309      & HBL  & 0.165      &758  &44.5   &$3.1\pm0.2$ & $^s$\\
1ES~1218+304    & HBL  & 0.182      &855  &45.2   &$3.0\pm0.4$ & $^t$\\
1ES~0347-121    & HBL  & 0.185      &864  &44.8   &$3.1\pm0.3$ & $^u$\\
1ES~1101-232    & HBL  & 0.186      &877  &44.8   &$2.9\pm0.2$ & $^v$\\
1ES~1011+496    & HBL  & 0.212      &1008 &45.5   &$4.0\pm0.5$ & $^w$\\
PG~1553+113     & HBL  & $>0.3$     &$>1498$& $>46.1$ & $4.5\pm0.3$ & $^x$  \\
                &      & $<0.74$    &$<4437$& $<47.1$ &             & $^x$ \\
S5 0716+714     & IBL  & 0.31   &1557 &46.1  &assumed 3.8  & $^{XY}$\\
3C279           & FSRQ & 0.536      &2998   &46.7     &  $4.1\pm0.7$  & $^y$   \\
\hline
\end{tabular}
\end{center}
\end{minipage}
\end{table}

\paragraph{M87} 
Besides the collective study of a Blazar sample \citep{2008MNRAS.385..119W}, it is interesting to closely examine nearby FR-I radio galaxies
as these objects are considered to be mis-aligned Blazars. The closest 
FR-I objects are M87 and Cen-A. VHE gamma-ray
emission from M87 had been initially reported from the HEGRA group \citep{2003A&A...403L...1A} and later verified by the Whipple collaboration \citep{2004ApJ...610..156L}, as well as by the H.E.S.S. group \citep{2006Sci...314.1424A} and most recently with the VERITAS telescopes \citep{2008arXiv0802.1951A}.\\
While the initial sensitivity was only sufficient to detect a signal and coarsely constrain the energy spectrum, following observations mainly with
the H.E.S.S. telescopes have lead to a number of important discoveries: the energy spectrum of M87 extends to energies beyond 10~TeV and even more spectacular,
the observed emission is highly variable on time-scales as short as days \citep{2006Sci...314.1424A}. The hard spectrum is a surprise and disfavors 
the models suggested by \citet{2005ApJ...634L..33G}
and \citet{2004A&A...419...89R} where the predicted gamma-ray spectrum is soft as a consequence of the assumption of a small Doppler-factor 
derived from radio measurements of the jet dynamics. It should be noted however,
that also other nearby  Blazars that have been observed with high resolution radio interferometers do not show strong super-luminal motion indicative
for large Doppler-factors. It seems
that there is growing evidence for a discrepancy between high Doppler factors implied by modelling the VHE emission of Blazars with single zone SSC model  
and  the lack of evidence for  strong super-luminal motion in these objects from radio observations \citep{2008arXiv0801.2749P}. 
\begin{figure}
\parbox{0.45\linewidth}{
\includegraphics[width=\linewidth]{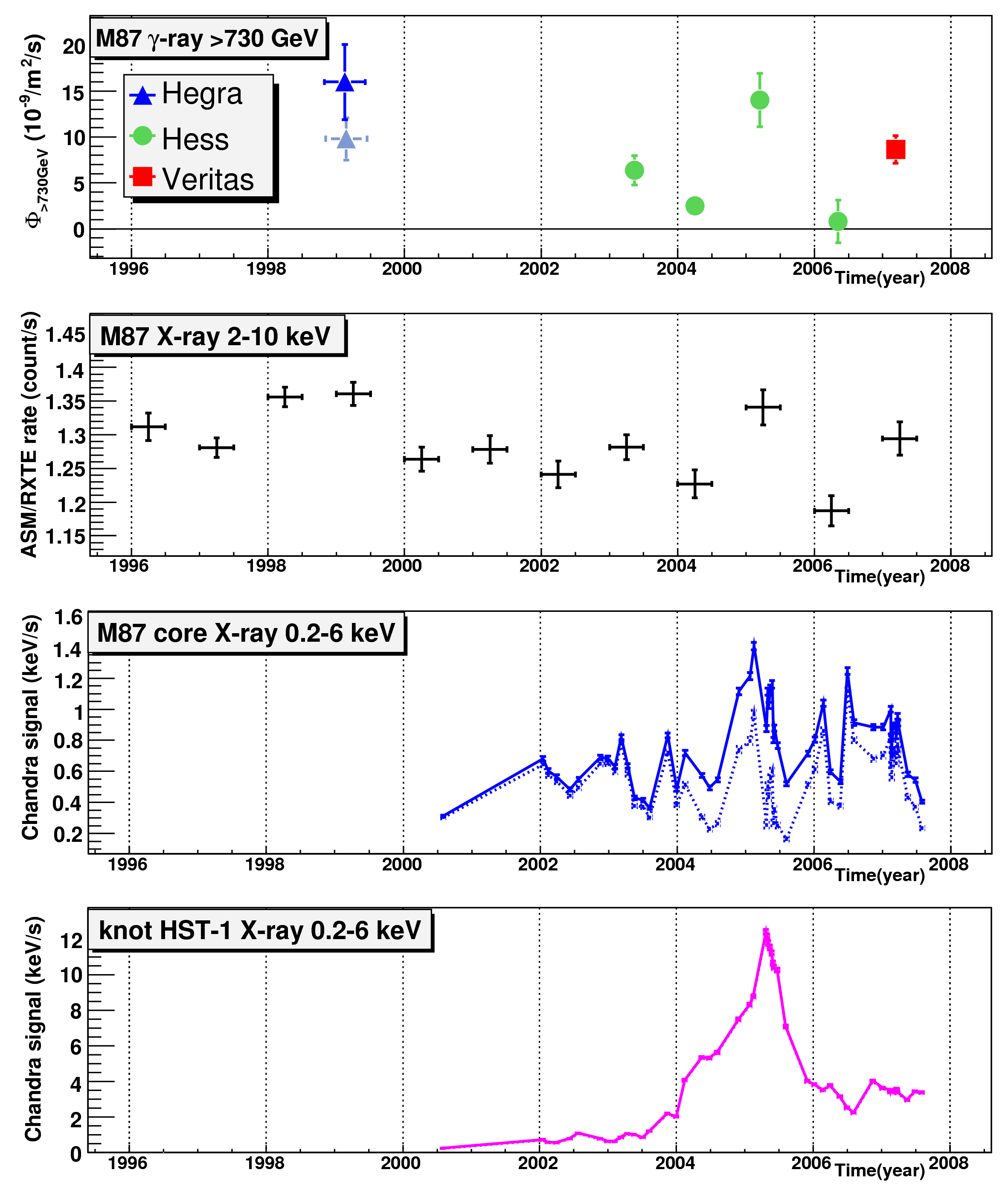}
}
\parbox{0.45\linewidth}{
\includegraphics[width=\linewidth]{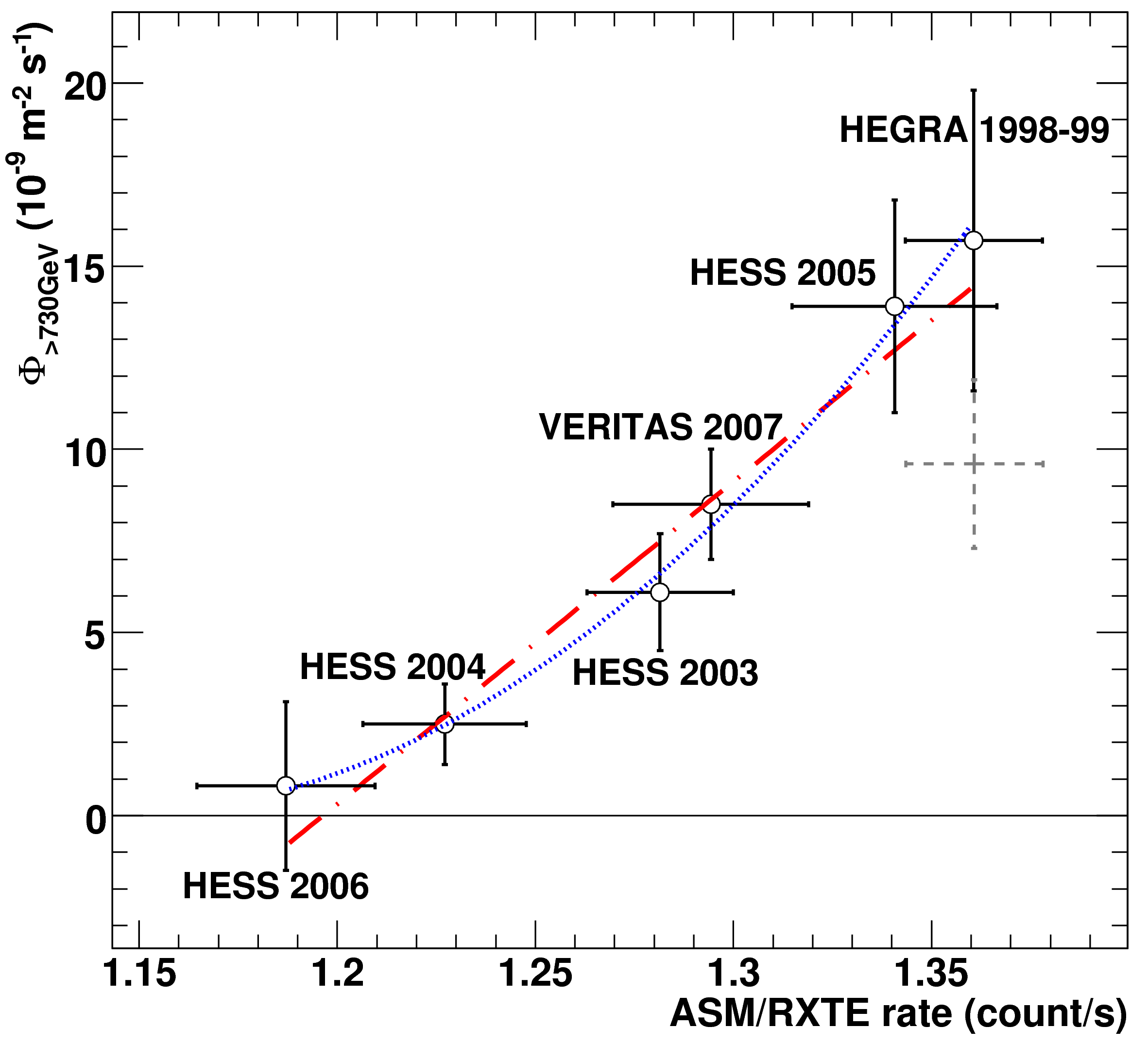}
}
\caption{\label{fig:m87} From \citet{2008arXiv0802.1951A}: The left panel combines the VHE and X-ray 
light curves from observations of M87 with ground based Cherenkov telescopes as well as with satellite instruments on-board the RXTE satellite (ASM) and the Chandra observatory. The right panel correlates the yearly
VHE fluxes with the appropriately averaged ASM fluxes. The dashed curve is a linear correlation
and the solid curve a squared function. }
\end{figure}
The fast variability observed from M87 leads to interesting implications. First of all, the variability excludes models where the VHE gamma-rays are
produced in processes like Cosmic ray interactions in the host galaxy \citep{2003A&A...407L..73P} or even Dark Matter annihilation \citep{2000PhRvD..61b3514B}. These processes should not lead to 
temporal variation of the observed emission. Given that the variability time scale can not be smaller than the light crossing time, the constraint  on the size of the emission region is severe
and excludes regions in the outer radio jet as well. A possible candidate for the emission could be the notoriously variable
HST-1 knot located less than 1 arc sec off the core. The most recent long-term light-curve obtained with the all-sky-monitor on-board the RXTE satellite as well as a dedicated Chandra monitoring seems
to support a correlation of the HST-1 activity in X-rays and the observed VHE emission (see Fig.~\ref{fig:m87}). 
\paragraph{3C279} This object belongs to the so-called flat-spectrum radio quasars (FSRQs). These objects are long-known to be emitters of GeV gamma-rays 
that have been detected with EGRET \citep{1992ApJ...385L...1H,2001ApJ...553..683H}. There are two aspects which single out this source: (a) 3C279 has a comparably large
red-shift of $z=0.536$ and (b) it is a new type of AGN which shows acceleration to TeV energies. \\
The EGRET Blazars are mostly members of the FSRQ-class and are characterized by a spectral energy distribution 
with a peak in the EGRET energy range, dominating the
total luminosity. The observed emission is commonly attributed to external Comptonization of photons from the broad-line region and possibly from
the accretion disk. Given that FSRQs have intrinsically a higher gamma-ray luminosity than Blazars, they can be in principle observed to higher red-shifts
opening up the possibility to finally detect absorption features in the VHE gamma-ray spectra (see also next section) of extra-galactic objects (see also
next section).  The detection of VHE emission from 3C279 with the MAGIC
telescope \citep{2007arXiv0709.1475T} indicates variability which seems to  be correlated with changes
in the optical flux. Further observations of this source and a measurement of the VHE energy spectrum could provide important 
clues on the level of extra-galactic background light (see next section). Even in the case of a 
low level of EBL, only $10^{-3}$ of the emitted flux will be observable at 1~TeV \citep{2008arXiv0802.0129R}. 
 \section{"Secondary physics": Propagation of gamma-rays}
\label{sec:secphy}
\subsection{EBL absorption}
VHE emission from AGN appears to be more common than what could have been hoped for and the temporal variability observed at the highest energy
is providing us an important and unique in-sight into the ``engine-room'' of AGN. The understanding of processes leading
to VHE-emission hinges however on a good understanding of pair-absorption processes of the type $\gamma_\mathrm{VHE}\gamma_\mathrm{EBL} \rightarrow e^+e^-$. 
This process has been known for a long time to be relevant for the propagation of TeV-photons over cosmological distances \citep{1966PhRvL..16..252G}.\\
With the growing number of Blazars observed at various red-shifts it has become feasible to use VHE-photons as \textit{probes} to constrain or even measure
the background photon density in extra-galactic space. Until now, it has however not been possible to clearly 
establish absorption in the measured Blazar spectra because the intrinsic emission spectrum
is not very well known and may suffer additional internal absorption \citep{2008arXiv0801.3198A}. When invoking a constraint on the
shape of the intrinsic spectrum, it is however possible to derive (model-dependent) upper limits on the EBL density \citep{2005ApJ...618..657D,2006Natur.440.1018A,2007A&A...471..439M}. The situation will be greatly improved when the first broad-band spectra of Blazars will be
obtained that should show a transition from the optically thin to the optically thick part. The detection of such an absorption feature will
 in turn lead to a detection of the EBL. In combination with increasing statistics
and coverage over varying red-shift, it will be feasible to investigate the evolution of the EBL and finally, 
derive in a completely independent way constraints for cosmological parameters \citep{2005APh....23..598B,2005APh....23..608B}.
\subsection{Lorentz invariance violation}
 VHE photons are useful probes for Lorentz invariance violating (LIV) processes. A number of theoretical approaches towards a quantum theory of Gravity predict a structure of space-time at the Planck scale. This will result in a modified dispersion relation for photons of the form $c^2p^2 =E^2(1+\xi(E/M_\mathrm{QG})+{\cal O} (E/M_\mathrm{QG})^{2}$) see e.g. \citet{2002MPLA...17.1025S} for a review. This dispersion relation affects the time-of-flight of photons of different energies and the kinematics of interaction,
suppressing e.g. the pair-production process of photons.\\ Currently, fast flares from Mkn~421 \citep{1996Natur.383..319G}, Mkn~501 \citep{2007ApJ...669..862A}, and PKS~2155-304 \citep{2007ApJ...664L..71A} have been observed. The latter is potentially the most constraining
observation as the red-shift is larger and the variability faster as well as more photons have been collected than for the other sources.
 Current limits on $M_\mathrm{QG}>3\times 10^{17}$~GeV \citep{2007arXiv0708.2889A} are already an order of magnitude better  
than limits obtained from GRBs \citep{2003A&A...402..409E,2006APh....25..402E}. It is important to point out that if a dispersion of 
arrival times from flares is detected, source intrinsic spectral variations can be excluded once consistent
time lags are detected from a sample of flares from different sources at different red-shifts. Ultimately,
the inferred LIV characteristics should be consistent with a suppression of EBL absorption processes. 
\section{Concluding remarks on the future perspectives of the field}
The observational field of VHE gamma-ray astrophysics has been successfully
driven in the last years by ground based imaging air Cherenkov telescopes.  The
``high energy frontier'' of Astrophysics will be expanded by the results
expected from the Fermi satellite.  The all-sky sensitivity of the Fermi
Large-Area-Telescope (LAT)  will most likely lead to many interesting new
discoveries and surprises. The inter-play of the Fermi-LAT detections from
space and follow-up observations from ground will provide mutual benefits for
the communities as well as  certainly answer many  questions as well as
stimulate new ones.  \\ While space-based observations of gamma-rays will be
for a long time  limited to the results obtained with Fermi (until maybe pair
conversion telescopes will be deployed on the surface of the moon), ground
based observatories will continue to be developed and constructed during the
next decade(s). To my knowledge, besides Fermi, no further space-based
gamma-ray telescope, operating in the GeV energy range, is planned.
Consequently, ground based gamma-ray detection techniques will be the only
available experimental approach to observe high energy gamma-rays after the
termination of the Fermi mission.\\ The extensions of H.E.S.S. and MAGIC into
phase II until 2009 will lower the energy threshold and improve existing
sensitivity moderately.  The next generation of ground based installations will
become fully operational at the end of the next decade as envisaged in the
AGIS\footnote{Advanced gamma ray imaging system}\citep{2007arXiv0709.0704K} and
CTA\footnote{Cherenkov telescope array} \citep{2007AN....328..600H} projects.
These installations
 aim at an improvement in sensitivity by a factor of 10
and a widened reach in energy. At that time, the new ground based instruments
including larger ($>$km$^2$) installations like TenTen \citep{2008NIMPA.588...48R} will explore new energy windows above $\approx 10$~GeV as well as above 10~TeV.  \\
 However, there still remains a ``blind spot'':  the transient sky above
10~GeV will neither be explored very well with Fermi (limited photon rate) nor
with ground based instruments of the current generation (limited field of view
and energy threshold). New installations like HAWC \citep{2005AIPC..745..234S}
will improve the situation in the energy range above a few 100~GeV. The lesson
from the detection of fast transient events as seen from Cyg~X-1 and
PKS~2155-304 however is that we are very likely missing the fast variability of
XRBs and AGN. Proper coverage of these objects requires instruments with a
large field of view and  an energy threshold well below 100~GeV.

\label{section:future}

{\small
\bibliographystyle{apj}
\setstretch{0.8}
\bibliography{references}

}

\vfill

\end{document}